\documentclass[default]{sn-jnl}

\usepackage{graphicx}%
\usepackage{multirow}%
\usepackage{amsmath,amssymb,amsfonts}%
\usepackage{amsthm}%
\usepackage{mathrsfs}%
\usepackage{enumitem}
\usepackage[title]{appendix}%
\usepackage{xcolor}%
\usepackage{textcomp}%
\usepackage{manyfoot}%
\usepackage{booktabs}%
\usepackage{algorithm}%
\usepackage{algorithmicx}%
\usepackage{algpseudocode}%
\usepackage{listings}%

\theoremstyle{thmstyleone}%
\theoremstyle{thmstyletwo}%

\theoremstyle{thmstylethree}%

\begin{document}

\title[Article Title]{Electrical control of valley polarized charged biexcitons in monolayer WS$_2$}


\author*[]{\fnm{Sarthak} \sur{Das}}\email{ds.sarthak.92@gmail.com}
\author[]{\fnm{Ding} \sur{Huang}}
\author[]{\fnm{Ivan} \sur{Verzhbitskiy}}
\author[]{\fnm{Zi-En} \sur{Ooi}}
\author[]{\fnm{Chit Siong} \sur{Lau}}
\author[]{\fnm{Rainer} \sur{Lee}}
\author[]{\fnm{Calvin Pei Yu} \sur{Wong}}
\author*[]{\fnm{Kuan Eng Johnson} \sur{Goh}}\email{kejgoh@yahoo.com}



\affil[]{\orgdiv{Institute of Materials Research and
Engineering (IMRE)}, \orgname{Agency for Science, Technology and
Research (A*STAR)}, \orgaddress{\city{Singapore}, \postcode{138634}, \country{Republic of Singapore}}}






\abstract{Excitons are key to the optoelectronic applications of van der Waals semiconductors with the potential for versatile on-demand tuning of properties. Yet, their electrical manipulation is complicated by their inherent charge neutrality and the additional loss channels induced by electrical doping. We demonstrate the dynamic control of valley polarization in charged biexciton (quinton) states of monolayer tungsten disulfide, achieving up to a sixfold increase in the degree of circular polarization under off-resonant excitation. In contrast to the weak direct tuning of excitons typically observed using electrical gating, the quinton photoluminescence remains stable, even with increased scattering from electron doping. By exciting at the exciton resonances, we observed the reproducible non-monotonic switching of the charged state population as the electron doping is varied under gate bias, indicating a coherent interplay between neutral and charged exciton states.}

\keywords{valley polarization, trion and quinton, motional narrowing, exchange interaction, resonant excitation, coherent coupling.}



\maketitle

\section*{Introduction}\label{sec1}
Excitons in monolayer transition metal dichalcogenides (TMDs) have high binding energy \cite{chernikov2014exciton, van2018coulomb, das2019layer}, enabling strong resonances even at room temperature. This makes TMDs highly attractive for a wide range of applications in optoelectronics \cite{xiao2012coupled, xu2014spin, schaibley2016valleytronics, mak2018light} and quantum technologies \cite{bussolotti2018roadmap, goh2020toward, bussolotti2023valley}. Particularly, the valley-dependent optical selection rules in monolayer TMDs, due to the presence of two inequivalent valleys with spin–valley locking and a direct bandgap, allow high degree of optical control over valley polarizations \cite{mak2012control, yu2014valley, yan2015valley, yu2016valley} and valley coherence \cite{yang2015long, hao2016direct}.

Although neutral excitons can host highly polarized optical states \cite{mak2012control, yu2014valley, yan2015valley, yu2016valley, yang2015long, hao2016direct}, their charge neutrality makes excitons insensitive to electrical bias, limiting the on-demand control of excitonic properties, such as the resonance energies, radiative and non-radiative lifetimes, and photoluminescence (PL) brightness. Moreover, these states are susceptible to environmental noise, which can result in a loss of quantum information. The challenges are further exacerbated by interactions with charges, spins, disorders, and thermal excitations. As a result, an electrically controlled, valley-polarized photon source remains unfulfilled. 

Earlier studies have shown that the modulation of valley polarization of neutral and charged states can be implemented with gate voltage \cite{shinokita2019continuous, wu2021enhancement}, temperature \cite{wu2021enhancement}, screening \cite{gupta2023observation}, strain \cite{an2023strain} or magnetic field \cite{smolenski2016tuning} but simultaneous electrical control along with the polarized photon yield remains a challenge. Moreover, the majority of the strategies implemented to achieve external control with valley-polarized excitons compromise the photon yield due to different non-radiative losses. To address the limitations of neutral exciton and explore valley-polarized charge states, we utilize optically generated valley-polarized emission from the charged states, controlled via electrical injection.

In addition to the excitons, the monolayer TMDs can host a nexus of many-body quasiparticles such as three-particle charged excitons or trions \cite{mak2013tightly, singh2014coherent, singh2016trion, hao2016coherent, robert2021spin, zinkiewicz2021excitonic}, four-particle biexciton \cite{steinhoff2018biexciton, conway2022direct, muir2022interactions}, five-particle charged biexcitons or quintons \cite{you2015observation, mostaani2017diffusion, barbone2018charge, ye2018efficient, rodin2020collective, chatterjee2021probing, mostaani2023charge, das2022highly, wei2023charged} and even higher many body species \cite{van2022six,choi2023emergence}. The experimental observation of all these multi-particle complex states has its genesis in the high binding energy of the underlying excitons.

In the presence of free carriers, neutral excitons can capture additional carriers to form charged excitons. While these charged species are electrically controllable, it is important to ascertain that they still preserve the spin-valley coupling characteristics (shown by valley polarization). Here, we demonstrate dynamic control over electrically tuneable steady-state valley polarization, effectively preserving valley information in the presence of enhanced scattering. By resonantly exciting excitons, we can achieve a controlled generation of charged states, wherein multi-particle excitonic quasiparticles become coherently coupled to excitons. Investigating the coherent coupling between neutral and charged states through gate-dependent photoluminescence spectroscopy under excitonic resonance reveals potential mechanisms to control the quantum states further.

Harnessing valley selective carrier populations from helicity-controlled excitations holds immense potential for generating, storing, processing, and transmitting information based on valley state. The TMD is advantageous due to its ability to leverage both optical (direct bandgap) and electronic (quasi-particle bandgap with strong spin-orbit interactions) benefits. In addition, the electronic degree of freedom allows for ease of control and integration with conventional electronic systems. However, the robust electrical control of a valley state via quinton species has not yet been demonstrated. For the charged excitonic species, we demonstrate that valley polarization can be electrically switched between high- and low-polarized states along with its dynamic modulation. This marks a significant step towards achieving practical applications of valleytronic devices. We discuss the implications of these findings in the context of coherent control of charged excitons where the charged exciton population can be controlled with neutral exciton resonance along with the retention of valley polarization. Our findings not only shed light on the manipulation of valley-polarized charge states in monolayer TMDs but also underline their potential as robust platforms for preserving quantum information in the presence of various environmental disturbances.

\section*{Results and Discussions}\label{sec2}

\begin{figure}[h]%
\centering
\includegraphics[width=1.0\textwidth]{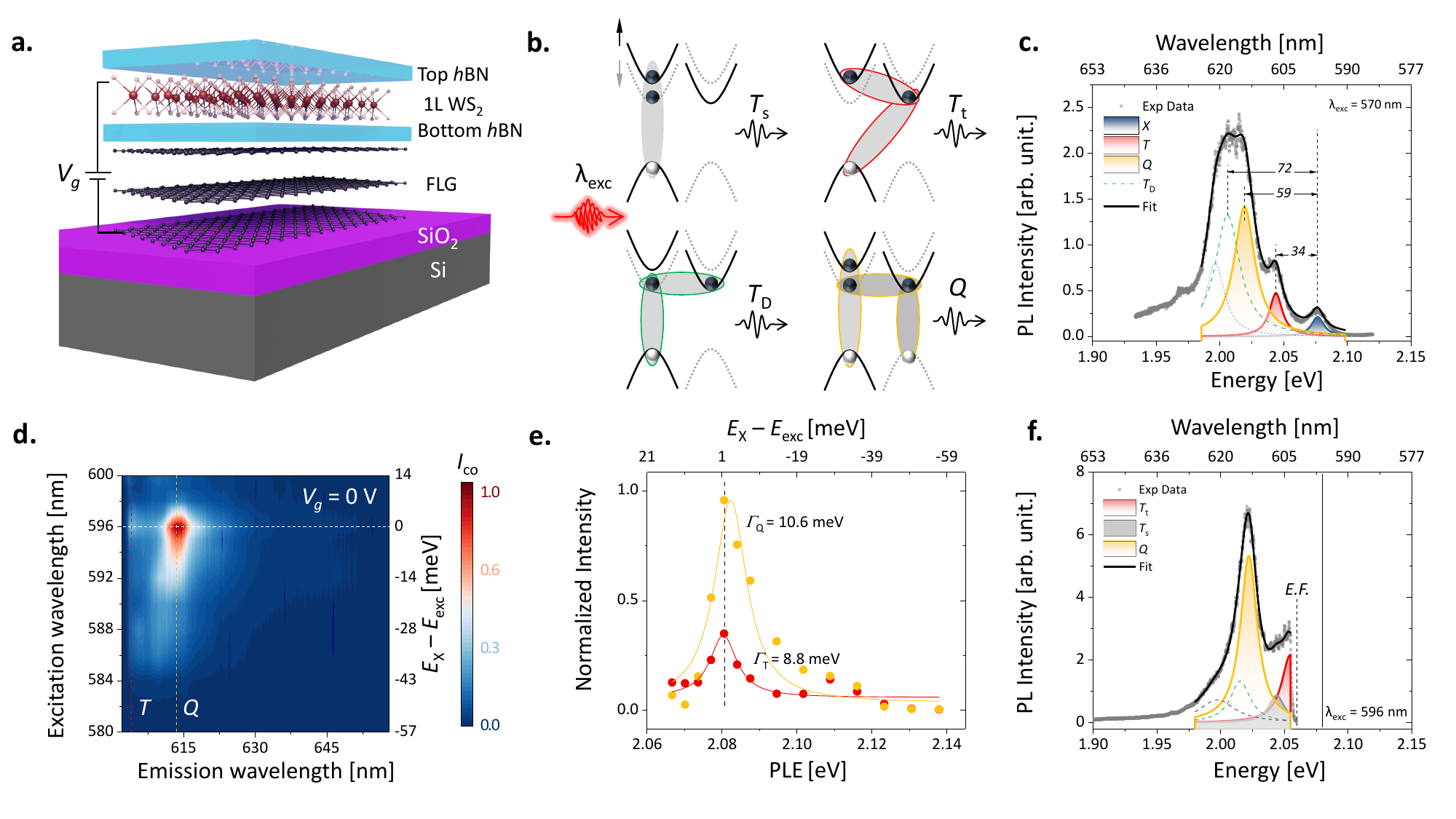}
\caption{\textbf{Device schematic and $\mu$PL of the excitonic complexes for resonant and non-resonant excitations.} 
(a) The schematic diagram of the sample device where monolayer WS$_2$ is sandwiched between two $h$BN layers over the graphite back gate. 
(b) The spin-valley configurations of $e-h$ pair for intravalley (singlet trion, $T_s$), intervalley (triplet trion, $T_t$), dark trion ($T_D$) and the combination of bright exciton ($X$) and $T_D$ or the five-particle charged biexcitons (quintons, $Q$). The solid (dotted) line represents up(-down) spin bands at the zone edges ($K$ and $K^\prime$ valleys). 
(c) Representative micro-photoluminescence ($\mu$PL) spectra of the monolayer for non-resonant $\lambda_{exc}$ = 570 nm excitation at $T$ = 4.6 K. Individual energy states are identified ($X$ at 596 nm or $\sim$ 2.078 eV), $T$ at 606 nm or $\sim$ 2.044 eV, $Q$ at 614 nm or $\sim$ 2.019 eV and $T_D$ at 618 nm or $\sim$ 2.006 eV) with multi-peak fitting. The energy separation relative to the neutral exciton (in meV) is marked for gate voltage, $V_g$ = 0 V with the excitation power $\sim$ 20 $\mu$W.
(d) The normalized intensity map of the excitation wavelength-dependent emission spectra recorded below the exciton energy at $T$ = 4.6 K. The right vertical axis represents the excess energy [energy difference between laser excitation ($E_{exc}$) and neutral exciton ($E_{X}$)]. The red (yellow) vertical dashed line denotes the $T$ ($Q$) energy resonance across the excitation. The brightest region corresponds to the exciton resonance ($\lambda_{exc}$ = 596 nm) denoted by the white horizontal dashed line with $V_g = 0$ V.
(e) The normalized intensity of the charged states (following the vertical lines in the previous figure) over the photo-luminescence excitation (PLE). The intensity is the maximum when the excess energy is the minimum. The profile is fitted with a Lorentzian curve to estimate the intrinsic broadening $\Gamma$ or carrier lifetime as $\tau = \hbar / 2\Gamma$.  
(f) Representative micro-photoluminescence ($\mu$PL) spectra of the monolayer for resonant $\lambda_{exc}$ = 596 nm excitation (following in horizontal line cut from figure d). The solid vertical line denotes the exciton resonance (excitation wavelength, $\lambda_{exc}$ = 596 nm) and the dashed line denotes the cut-off filter for reference. The individual peaks are fitted with Lorentzian multi-peak fitting to show the $T_s$, $T_t$ and $Q$ energy states with an incident laser power of $\sim$ 8.7 $\mu$W.}\label{fig1}
\end{figure}

Our sample consists of a monolayer of tungsten disulfide (1L WS$_2$) sandwiched between two hexagonal Boron Nitride ($h$BN) layers. A graphite layer underneath the bottom $h$BN layer serves as a back gate electrode (figure \ref{fig1}a). The 1L WS$_2$ is connected to an electrode via another graphite layer (see \textbf{Supporting Information S1} and \textbf{Methods} for further details). Just as a bright exciton in WS$_2$ can capture an electron and form intra- or inter-valley trions (singlet $T_s$ or triplet $T_t$ trion), a dark exciton can also form dark trion states ($T_D$) \cite{robert2021spin, zinkiewicz2021excitonic}. Also, due to the presence of the lowest energy dark states in WS$_2$, the bright exciton can combine with $T_D$ to form a five-particle quinton (or charged biexciton state, $Q$) \cite{barbone2018charge, ye2018efficient, chatterjee2022trion}. The different configurations of the energy states are schematically presented in figure \ref{fig1}b. The low temperature ($T$ = 4.6 K) micro-photoluminescence ($\mu$PL) spectra from the sample are presented in figure \ref{fig1}c with $\lambda_{exc}$ = 570 nm excitation (above optical bandgap, non-resonant) and back gate voltage $V_g$ = 0 V. The different energy states are extracted from the Lorentz fit and highlighted as exciton ($X$ at 596 nm or $\sim$ 2.078 eV), trion ($T$ at 606 nm or $\sim$ 2.044 eV), quinton ($Q$ at 614 nm or $\sim$ 2.019 eV) and dark trion ($T_D$ at 618 nm or $\sim$ 2.006 eV) based on their energy position according to the previous reports \cite{zinkiewicz2021excitonic, prando2021revealing, chand2023interaction}. Further experimental observations in support of the peak assignments are provided in \textbf{Supporting Information S2}.

The exciton population primarily governs the population of the coupled excitonic quasi-particles. Further, the exciton population can be controlled efficiently by resonant ($\lambda_{exc}$ = 596 nm) vs non-resonant excitation. Figure \ref{fig1}d shows a 2D color map of PL excitation (PLE) spectra measured near the neutral exciton resonance ($\sim$ 596 nm, horizontal dashed line) at $V_g$ = 0 V. At excitation energies higher than the neutral exciton emission, the spectra reflect an off-resonant regime where the PL emission of $T$ and $Q$ peaks is low due to fast non-radiative recombination processes \cite{das2022highly}. However, as the excitation energy approaches the exciton resonance ($E_X$), PL peaks of the individual energy states get stronger while the width is reduced (see the \textbf{Supporting Information S3}). The peak intensity reaches the maximum at the neutral exciton resonance (596 nm at $V_g$ = 0 V), then decreases due to sub-bandgap optical excitation. The red (yellow) vertical dashed lines follow the $T$ ($Q$) energy states. Figure \ref{fig1}e shows the normalized intensity profile at $T$ and $Q$ resonances as a function of excitation energy, highlighting the PL enhancement of the charged excitons (at the $X$ resonance shown by the black dashed line) due to efficient down-conversion from the resonantly formed neutral excitons. The Lorentzian profile broadening along the resonance energy gives an estimation of the intrinsic linewidth corresponding to the multi-particle complex states. This also denotes the efficient coupling between the neutral and charged states \cite{hao2016coherent, ayari2020phonon}. Figure \ref{fig1}f represents the horizontal line cut from figure \ref{fig1}d at exciton resonance with $\lambda_{exc}$ = 596 nm. The solid vertical line denotes the excitation wavelength at 596 nm, while the dashed vertical line represents the edge of the cut-off filter for reference (see the \textbf{Methods} section for further details). The different energy states ($T_t$, $T_s$, and $Q$) below the $X$ state are well identified for resonant excitation at $X$.

\begin{figure}[h]%
\centering
\includegraphics[width=1.0\textwidth]{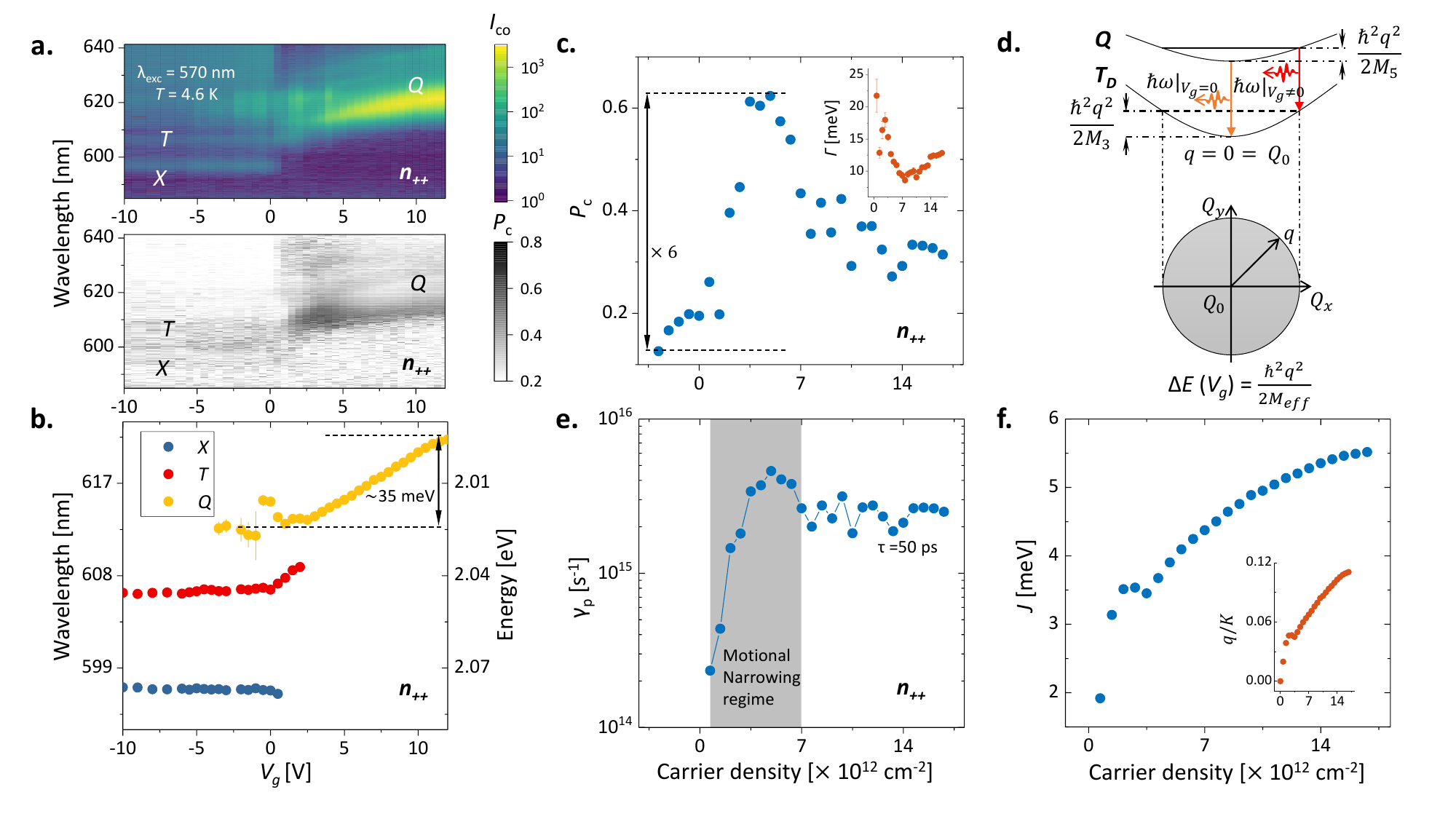}
\caption{\textbf{Gate dependent spectroscopy for non-resonant excitation and dynamic modulation of valley polarization for the charged states.} 
(a) The two-dimensional colorplot for the $V_g$ dependent $\mu$PL spectroscopy under non-resonant excitation shown in the upper panel with co-polarized mode (where $I_{co}$ is the intensity under $\sigma^+$ excitation and $\sigma^+$ detection, $\sigma^+$/$\sigma^+$). The energy states are mostly featureless for $V_g < 0$ V. However, when electron injection starts (for $V_g > 0$ V) the neutral state intensity drops and charged state intensity grows monotonically with a pronounced red shift dominated by the $Q$ state in the high $n$-doped region (labelled as $n_{++}$). The bottom panel represents the colorplot of the corresponding $P_c$ within the wavelength range as a function of gate voltage. 
(b) The gate-dependent variation in energy positions corresponding to different energy states extracted from the individual peak fittings. The $Q$ state shows a noticeable red shift under positive gate bias of $\sim$ 35 meV compared to the neutral region. 
(c) The gate dependent valley polarization $\mathcal{P_C}$ or the degree of circular polarization ($P_c$) of the $Q$ state. The $P_c$ shows a non-monotonic behavior as the carrier density increases. Inset: Corresponding linewidth ($\Gamma$) of the $Q$ state extracted from the fitting across the carrier density. 
(d) Representative diagram describing about the $V_g$ dependent transitions from charged biexcitonic state (quinton state, $Q$) to a $T_D$ state. At $V_g = 0$ V the vertical transition (represented by the orange arrow) is at $q = 0 = Q_0$ while the transition occurs ($\hbar\omega$ at $V_g > 0$ V) at higher $q$ value (represented by the red arrow) for higher $V_g$ because of Pauli blocking. In-plane projection of the same is presented in the bottom panel in the $Q_x-Q_y$ plane for reference. 
(e) The extracted scattering rate, $\gamma_p$, for a given lifetime of $\tau$ = 50 ps for $Q$ state \cite{you2015observation, barbone2018charge, chatterjee2021probing, das2022highly}. The scattering rate increases initially along with the increase in carrier density and then saturates. The increment in $P_c$ along with $\gamma_p$ refers to the motional narrowing regime (shaded region).
(f) The extracted exchange interaction ($J^{LR} (q)$) variation with carrier density along with the normalized CoM momenta dispersion ($q/K$) mapped from the kinetic energy difference (inset) for the experimental red shift from figure \ref{fig2}b using equation \ref{delE}. The $J^{LR} (q)$ increases with increasing carrier density.}\label{fig2}
\end{figure}

When an external DC bias is applied to the monolayer via the back gate, as we can observe in figures \ref{fig2}a-b, the energy states exhibit a shift in energy position, change in intensity, and variation in the valley polarization as reported in other works \cite{barbone2018charge, chatterjee2022trion, das2022highly}. The two-dimensional colorplot for gate-dependent PL spectra for non-resonant excitation ($\lambda_{exc}$ = 570 nm) is presented in the upper panel of figure \ref{fig2}a. The representative spectra and fittings are shown in \textbf{Supporting Information S4}. Due to the optical selectivity and unique spin-valley configuration, the right (left) circularly-polarized light can selectively address the $K (K^\prime)$ valley in 1L WS$_2$. Here, the PL intensity is recorded for both the co-($\sigma^+$ excitation and $\sigma^+$ detection, $\sigma^+/\sigma^+$) and cross-polarized ($\sigma^+$ excitation and $\sigma^-$ detection, $\sigma^+/\sigma^-$) modes (see \textbf{Methods} section for further details). The intensity mapping presented in figure \ref{fig2}a is for the co-polarized mode, while the cross-polarized spectra are shown in \textbf{Supporting Information S5}. The corresponding degree of circular polarization ($\mathcal{P_C}$) in steady-state can be obtained from the relation:
\begin{equation}\label{DoCP}
\mathcal{P_C} = \frac{I_{co} - I_{cross}}{I_{co} + I_{cross}}   
\end{equation}
where $I_{co/cross}$ denotes the intensity of the co- or cross-polarized PL. All the energy states discussed above show a distinct polarization upon circularly polarized light excitation. The bottom panel in \ref{fig2}a, shows the color plot of $\mathcal{P_C}$ as a function of applied $V_g$. The peak positions of the individual energy states are extracted from the individual peak fittings and plotted in figure \ref{fig2}b. We observe three regions with distinct behaviors as follows:

(i) For $V_g < 0$, the individual energy state's position and corresponding intensity remain unaltered. The charged excitonic complex states are highly sensitive to doping. However, due to the relative band alignment between 1L WS$_2$ and graphite, it is difficult to inject holes into the WS$_2$ \cite{lau2020quantum, lau2022gate}. Hence, the exciton and other excitonic complexes undergo negligible modulation while $V_g$ is negative.

(ii) At small positive $V_g$, the electrons are injected into the system. These free electrons are captured by the optically created neutral excitons ($X$) and thus form three-particle negatively charged exciton ($T$) or five-particle quintons ($Q$, charged biexciton). Also, in this process, neutral excitons transfer their oscillator strength to the charged species. As a result, the intensities of the $T$ and $Q$ states start to increase (see \textbf{Supporting Information S5}) while the $X$ intensity fades away.

(iii) As doping increases further ($V_g > 0$), the spectrum is dominated by the five-particle $Q$ state. This is because the lowest energy state in 1L WS$_2$ is dark \cite{liu2013three, zhang2015experimental, echeverry2016splitting}, thus favoring the formation of $Q$ states rather than $T$ states. This $Q$ state experiences monotonic enhancement of the intensity and gradual red-shift due to Pauli blocking with the increasing $V_g$ \cite{ye2018efficient, chatterjee2022trion} (marked as $n_{++}$ in the figures \ref{fig2}a-b). The red shift of the $Q$ state is almost linear over the experimental range, with a slope of $\sim$ 35 meV per decade.

The $\mathcal{P_C}$ corresponding to the $Q$ state is presented in figure \ref{fig2}c. As shown in figures \ref{fig2}a-b, the energy states are not sensitive to the external bias for the $V_g < 0$ V region. This is also true for $\mathcal{P_C}$. As the intensity of the $Q$ state grows with the number of free electrons, the $\mathcal{P_C}$ of the Q state is modified non-monotonically. As shown in figure \ref{fig2}c, $\mathcal{P_C}$ changes from $<$ 10\% to $>$ 60\% at carrier density ($n$) of $\sim$ 4 $\times$ $10^{12}$ cm$^{-2}$ (six-fold enhancement) then gradually decreases and finally saturates at $\sim$ 30 \% after $\sim$ 8 $\times$ $10^{12}$ cm$^{-2}$. Earlier studies have shown that the modulation of $\mathcal{P_C}$ of neutral ($X$) and charged ($T$) states can be implemented with gate voltage \cite{shinokita2019continuous, wu2021enhancement}, temperature \cite{wu2021enhancement}, screening \cite{gupta2023observation}, strain \cite{an2023strain} or magnetic field \cite{smolenski2016tuning}.  Importantly, the modulation of the neutral states by electrostatic doping is limited to the low carrier density regime only. Hence, the dynamic modulation of $\mathcal{P_C}$ in charged states is of particular interest.

Since the $n$ doping is increasing linearly with the gate voltage, the non-monotonic behavior of $\mathcal{P_C}$ can be understood as a product of parallel contributions from different depolarization mechanisms. Here, the non-monotonic behavior is explained using a combination of factors such as long-range electron-hole exchange ($J^{LR}$), $V_g$-dependent dispersion of quinton state in exciton momentum ($q$) space, and the corresponding momentum scattering rate ($\gamma_p$). Recently, the Maialle-Silva-Sham (MSS) mechanism caused by the intervalley electron-hole ($e–h$) exchange interaction has been suggested to dominate the spin and valley relaxation in monolayers \cite{yu2014dirac, hao2016direct, chen2018superior, gupta2023observation}. Here, the $J^{LR} (q)$ acts as a momentum-dependent effective magnetic field where different multi-particle energy species with a different center of mass (CoM) precess around the effective magnetic field with different frequencies \cite{yan2015valley, gupta2023observation}. The exchange-driven steady-state valley polarization can be expressed as \cite{chen2018superior,wu2021enhancement}:
\begin{equation}\label{Pc}
    \mathcal{P_C} = \frac{1}{1+\frac{J^{LR}(q)}{\hbar/\tau[\hbar/\tau+\hbar/\tau_p]}}
\end{equation}
Here $\tau$ refers to the radiative lifetime of the state, $\tau_p$ is the depolarization related to momentum scattering, with $\frac{1}{\tau_p}$ = $\gamma_p$ being the momentum scattering rate, while the intervalley exchange interaction with CoM momentum is $q$, and $J^{LR}(q)$ reads as 
\begin{equation}\label{J}
 J^{LR}(q) = - |\psi (r_{e-h} = 0)|^2 (\frac{at}{E_g})^2 V_q q^2 e^{-2i\theta}   
\end{equation}
where $|\psi (r_{e-h} = 0)|^2$ is the overlap between electron and hole wavefunction which is approximated by $1/{a^2_B}$, $a_B$ refers to the Bohr radius, $a$ is lattice constant, $t$ is the hopping parameter, $E_g$ is the electronic bandgap, and $V_q$ is the Coulomb potential defined by $V_q = \frac{2 \pi e^2}{\epsilon q} \frac{1}{1+r_0q}$, where $e$ is the elementary charge, $\epsilon$ is the effective dielectric, and $r_0$ is the fitting parameter. 

For recombination of the neutral excitons, the radiative photon is confined within a light cone due to the conservation of momentum, resulting in a small kinetic energy. In contrast, the constraint on the recombination of charged states is not as stringent, as the energy of the final state for recombination of the charged state changes with $q$ \cite{wang2016radiative, chatterjee2022trion}. Thus, a recombination process of charged states can also occur at different $q$ values. While the recombination at $V_g$ $\approx$ 0 V is from the band edge $q = 0$, the recombination occurs at higher $q$ values when $V_g$ increases. Now, the difference in effective mass between the five-particle $Q$ state and the three-particle $T_D$ state results in pronounced kinetic energy. To conserve energy and momentum, the emitted photon undergoes a redshift over the applied voltage range as shown in figure \ref{fig2}d. The energy difference can be mapped from the following relation:
\begin{equation}\label{delE}
    \Delta E (V_g) = \hbar\omega (V_g) - \hbar\omega (V_g = 0) = \frac{\hbar^2 q^2}{2} [\frac{1}{M_3} - \frac{1}{M_5}]
\end{equation}
Here $M_3 (M_5)$ is the mass of the three-(five-) particle species, and $\hbar\omega (V_g)$ denotes the photon energy for given $V_g$. As the carrier density or $n$ doping increases, the $J^{LR} (q)$ also increases due to $q$-space dispersion. The $V_g$-dependent $J^{LR} (q)$ can be used in equation \ref{Pc} to match the experimental $\mathcal{P_C}$ value for a given $\tau$. Further details on this are described in \textbf{Supporting Information S6}. Assuming the radiative lifetime is constant across $n$, we can further extract the momentum scattering rate ($\gamma_p$).

In figure \ref{fig2}e, $\gamma_p$ gradually increases with carrier density at the low-doping regime and then saturates after $n$ reaches $\sim$ 4 $\times$ $10^{12}$ cm$^{-2}$. If the $\tau$ of the charged state is constant over the carrier density, we observe a strong variation in $\gamma_p$ for the charged excitons, owing to their increased $q$-space dispersion. The initial increase in momentum scattering leads to the enhancement of valley polarization observed in figure \ref{fig2}c. Similar behavior was recently reported for monolayer TMDs, where the spin/valley polarization of neutral exciton increases with an increase in carrier density \cite{yan2015valley, wu2021enhancement}, and this is attributed to the motional narrowing effect \cite{dal2015ultrafast, cadiz2017excitonic}. However, with higher doping, this effect vanishes with enhanced kinetic energy and $J^{LR} (q)$ (see \textbf{Supporting Information S6}). This is also reflected in the width reduction of the PL peaks at the low-doping regime, as shown in the inset of figure \ref{fig2}c and \textbf{Supporting Information S5}. Contrary to the almost constant, small value of the exchange interaction for the neutral exciton, the $J^{LR} (q)$ of the charged states show pronounced variation \cite{hao2016direct, chen2018superior, wu2021enhancement} with increasing $n$ as shown in figure \ref{fig2}f. Also, the CoM momentum distribution of the $Q$ state across the carrier density, extracted from the experimental red-shift of the peak position (from figure \ref{fig2}b and equation \ref{delE}) presented in the inset of figure \ref{fig2}f.

\begin{figure}[h]%
\centering
\includegraphics[width=1.0\textwidth]{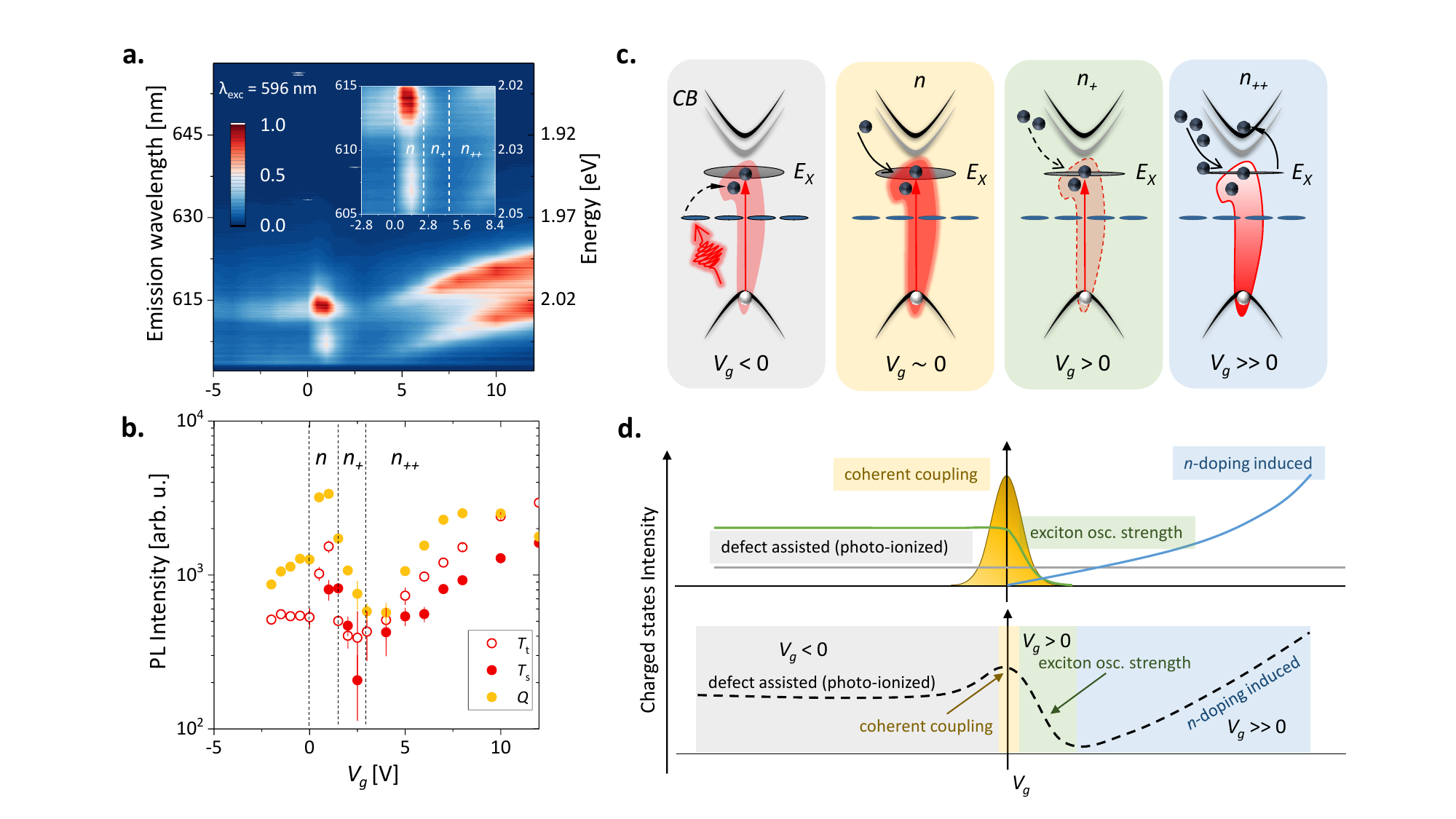}
\caption{\textbf{Modulation of coherently coupled charged excitonic complexes under resonant excitation.} 
(a-b) Gate-dependent intensity mapping of the charged states under resonant excitation with circularly polarized light in co-polarized ($I_{co}$ or $\sigma^+/\sigma^+$) mode. A zoomed-in picture is presented in the figure inset divided into four different regions. The intensity profile follows a non-monotonic behavior. Non-monotonic intensity modulation of the charged states corresponding to $T_s$, $T_t$ and $Q$ states is presented in b. The vertical dashed lines demarcate four different regions same as in a. 
(c) Formation of the charged states are represented schematically for four different conditions (Three particle states are shown in the schematics for simplicity. The same argument holds for five particles also). For $V_g < 0$ V (gray background), the charged states are formed either through defect/impurity-assisted processes or through photo-ionization effect (defect states are shown by the discontinuous blue band below the optical bandgap and the neutral exciton energy is denoted as $E_X$ with the oscillator strength represented by the width of the $E_X$ band). This is a relatively weak and inefficient process, so the population of charged states is low. When $V_g$ becomes slightly positive, the resonantly formed excitons (red vertical arrow) can efficiently capture the electrically injected (shown by the curved arrow) carriers and hence there is a strong population of charged states ($n$: yellow background). For the $n_+$ region, the exciton oscillator strength drops with further $n$-doping. Since, the net population density of the charged states is a product of the exciton population (oscillator strength of the neutral exciton) and the coupling parameter under resonance, hence, the coherently coupled charged state density decreases ($n_+$: green background). Finally, for $V_g >> 0$ V, the system becomes electron-rich and the $n$-doping induced charged state carrier population starts to grow as for the non-resonant excitation scenario ($n_{++}$: sky background). 
(d) The upper panel schematically represents the individual expected intensity profile of the charged states as a function of $V_g$ for the four cases described above while the combined effect is represented in the bottom panel as a reference describing the anomalous experimental trend-line (with black dashed line). The color coding refers to the different cases for four different regions of $V_g$ described above.
}\label{fig3}
\end{figure}

As discussed earlier, the formation of the charged states under resonance can be controlled efficiently via coherent coupling between neutral and charged excitons \cite{singh2014coherent, hao2016coherent, kallatt2019interlayer}. The neutral exciton population is more dominant with resonant excitation, while the charged states population can be modulated via external doping. Hence, changing the carrier density at the exciton resonance opens up a channel to control the coupling parameters between neutral and charged states. The 2D colorplot in figure \ref{fig3}a represents the PL for the lower energy states relative to $X$ across the range of $V_g$ (see the \textbf{Methods} section for further details). Here, the excitation energy is set to neutral exciton resonance (596 nm). Unlike the non-resonant case, the intensity of the $T$ and $Q$ charged quasi-particles excited at the exciton resonance exhibit a non-monotonic gate dependence, as shown in figure \ref{fig3}b.

For $V_g < 0$ V, we do not observe evidence of significant external doping in figure \ref{fig3}a ($p$ doping is highly unlikely \cite{lau2020quantum, lau2022gate}), and the Fermi level stays within the bandgap. Hence, the exciton population remains unaltered throughout this region (as shown in figure \ref{fig3}a). Also, since the excitation is exactly at the excitation resonance, there is a scarcity of optically generated free carriers. Thus, the charged excitons are formed either through defect-assisted processes \cite{gao2016valley} or via photo-ionized carriers \cite{mitioglu2013optical}. As a consequence, there is no significant variation in the population of charged excitons, and their total density remains relatively low.

At small positive gate voltages ($0 < V_g < 1$ V), we observe in figures \ref{fig3}a-b that the population of free electrons starts to increase (denoted as $n$), promoting the formation of charged $T$ and $Q$ species. The population of charged excitons is further expected to be enhanced by the resonant condition that assists the down-conversion of the neutral excitons \cite{ayari2020phonon}. The resonantly formed excitons can now efficiently capture carriers and form multi-particle states. In this regime, the coherent coupling between neutral and charged excitons reaches the maximum, resulting in an order-of-magnitude increase of the charged exciton densities. Under resonance, the net population density of the charged states is a product of the exciton population (oscillator strength of the neutral exciton) and the coupling parameter. As the doping increases with increasing $V_g$, the exciton oscillator strength decreases, causing a decrease in the population of coupled charged excitons, reaching its minimum at $V_g = +2.5$ V (denoted as $n_+$ in figures \ref{fig3}a-b).

For $V_g > 3$ V, where the system is electron-rich in figures \ref{fig3}a-b, the steady-state population of the charged states is not coherently coupled. This can further be confirmed by performing the PLE experiment similar to figure \ref{fig1}d but in an electron-rich environment (shown in \textbf{Supporting Information S7}). As a consequence, it grows with increasing doping, similar to non-resonant excitation, as shown in figure \ref{fig2}a (denoted as $n_{++}$ region).

Here, the charged states are primarily dependent upon the neutral exciton population, especially for the low-doping regime. This anomalous behavior of the charged states population is a unique example of the doping dependence of the coherent coupling between neutral and charged states. The above plausible scenarios are schematically presented in figure \ref{fig3}c for four different regimes ($V_g < 0$ V, and three $n$-doping regions marked as $n$, $n_+$ and $n_{++}$ are presented with different background colors). Resonant excitation persists throughout the applied voltage and is shown via a red vertical arrow, and the relatively weaker processes are denoted by the dashed curved arrows. The intensity profile for the individual effects is demonstrated in the top panel of Figure \ref{fig3}d, and the collective trend is depicted in the bottom panel of the same figure (the color coding follows the same conventions as those in the prior figure). The intensity modulation of charged states with the change in gate voltage is due to the combination of the various processes described above.

Similar to the non-resonant regime, the motional narrowing effect for valley polarization of charged excitons persists with resonant excitation of neutral species at low doping densities ($<$ 4$\times$10$^{12}$ cm$^{-2}$) (see discussions in \textbf{Supporting Information S8-S9}). The efficient down-conversion from the neutral to the charged states can be enhanced in the presence of resonant optical phonons (Fr\"{o}hlich interaction) \cite{miller2019tuning}. Indeed, the energy difference between the $X$ and $T$ states matches quite well with the vibrational phonon mode ($E_{2g}^1$ mode) of the monolayer WS$_2$. Hence, the existence of phonon-driven ``virtual trion" is highly probable \cite{van2019virtual}. However, it has been reported that only optical phonons are favorable for this efficient down-conversion, and so the same argument does not hold for the efficient generation of the $Q$ state from the $X$ state resonance \cite{shree2018observation, van2019virtual}. While there is no optical phonon with energy that can match the energy difference between $Q$ and $X$ excitons, the down-conversion may result from interactions with overtones or a combination of (optical and acoustic) phonon modes \cite{molas2017raman} (see \textbf{Supporting Information S11} for further details).

\begin{figure}[h]%
\centering
\includegraphics[width=1.0\textwidth]{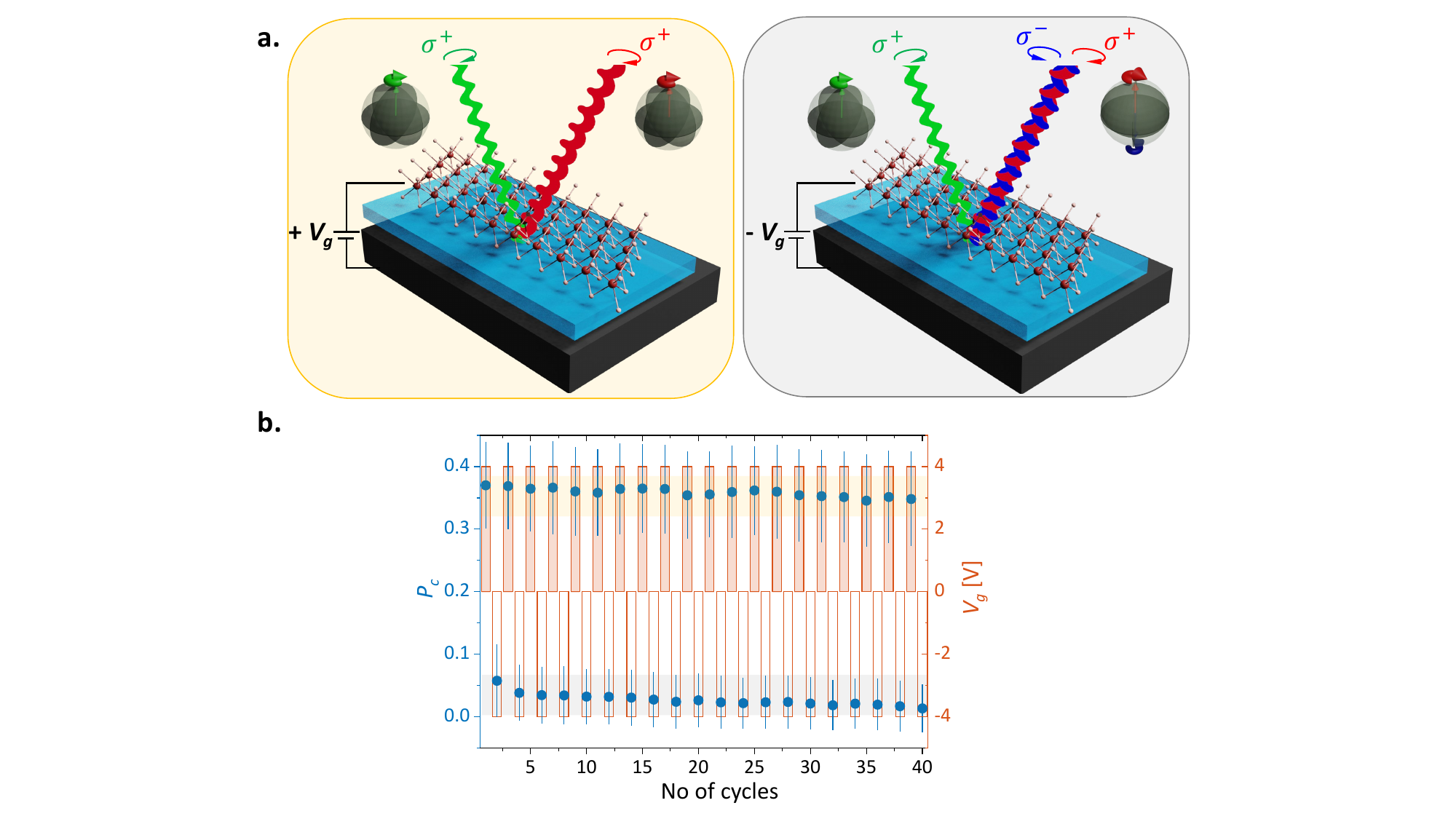}
\caption{\textbf{Electrical switching of polarization for quinton states.} 
(a) The degree of circular polarization corresponding to the quinton states can be electrically switched from a high-polarized state to a low-polarized state by altering the polarity of $V_g$. The circular polarization retains for the bright $Q$ states when $V_g$ is positive i.e. the system is electron-doped as shown in the left panel. However, the valley information is lost when the polarity of the $V_g$ is altered as shown in the right panel. In our experimental scheme, we excite with $\sigma^+$, and detect $\sigma^+$ photon (shown in red) from the $Q$ state with a positive voltage, while getting the equal intensity of both the helicity of photons with negative $V_g$ (shown in red and blue). 
(b) The device operation is demonstrated over several cycles by sequentially altering the polarity of $V_g$ and recording the corresponding $P_c$ over a wavelength window of 618 - 624 nm. Here, the $V_g$ is switched between $\pm$ 4 V and the $P_c$ is switched between high (yellow-shaded region) and low-polarized states (grey-shaded region) efficiently without any noticeable variations. The error comes from the statistical average within the specified wavelength range.   
}\label{fig4}
\end{figure}

Finally, we show the repeatable voltage-dependent switching of the degree of polarization for reconfigurable $Q$ states for up to 40 cycles. By altering the $V_g$ polarity, the $P_c$ can be switched from the high polarized state to a low polarized state as shown in figure \ref{fig4}a. To show the reproducibility of our observations we have repeated all the operations in a separate device D2 similarly fabricated (details are described in \textbf{Supporting Information S12}). For a demonstration of stable switching control, we chose a threshold of $<$ 10\% for the low state and $>$ 40\% for the high state, for both the non-resonant and resonant excitation. Here, we record the average $P_c$ with the statistical error over a selected voltage window primarily within the $Q$ energy state over the cycles by altering the polarity of $V_g$. The recorded $P_c$ of the second device shows stable switching in either high or low-polarized states within measurement errors. The electrical switching of valley polarization corresponding to the bright $Q$ states is represented in figure \ref{fig4}b in the main text with quasi-resonant excitation, while the same for the first device is shown in \textbf{Supporting Information S13}.  

\section*{Conclusion}\label{sec3}
In conclusion, our work demonstrates the ability to deterministically control the valley-polarized emission from charged exciton states in monolayer TMDs using an external electrical bias, thereby overcoming the limitations of neutral excitons. Through careful selection of scattering mechanisms, we can generate strong valley-polarised emission of a given chirality or suppress it on demand. We also found that this polarization is resilient against environmental noise fluctuations associated with our devices. By using electrical bias, we can efficiently vary the exchange interaction and scattering rates of charged states, accentuating the motional narrowing effect, and yielding a deterministic valley-polarized emission from charged states and simultaneously boasting remarkable optical brightness. Specifically, we show electrical control over charged biexciton (quinton) which thereby enabled the deterministic control of valley polarization in monolayer WS$_2$ devices where $P_c$ can be tuned from $<$ 10\% up to $>$ 60\%, representing a 6-fold increase under electrical bias. The exquisite electrical tunability of the spectral resonance of charged states promises coherent control of valley polarization through the intricate interplay between the charged and neutral exciton states, and the electrical control of coherent valley polarization bodes well for exploiting the valley degree of freedom for information storage and manipulation. This research represents a key step towards advanced optoelectronic devices and opens up new possibilities for applications such as valley-selective optical switching and valley-based quantum technologies with enhanced stability and reliability.

\backmatter

\bmhead{Supplementary Information}

This article contains supplementary information from S1 to S13.

\bmhead{Acknowledgments}

We acknowledge the useful discussion with Prof. Goki Eda. We acknowledge the funding support from the Agency for Science, Technology and Research \#21709 and C230917006. D.H. acknowledges the funding support from A*STAR project C222812022 and MTC YIRG M22K3c0105, C.S.L. acknowledges the funding support from A*STAR under its MTC YIRG grant No. M21K3c0124 and C.P.Y.W. acknowledges funding support from the A*STAR AME YIRG Grant No. A2084c0179. 

\bmhead{Competing Interests}

The authors declare no competing financial or non-financial interests.

\bmhead{Data Availability}

Data is available on reasonable request from the corresponding author.

\bmhead{Author Contribution}

S.D. and K.E.J.G. conceptualized the experiment. S.D. fabricated the device with the help of R.L., C.S.L. and C.P.Y.W. and performed the experiments. D.H., I.V. and Z.E.O. helped in optical measurements. S.D. conducted the data analysis with inputs from D.H., I.V., and K.E.J.G. All authors contributed to the manuscript writing.

\bibliographystyle{naturemag}
\bibliography{references}

\newpage

\end{document}


\title[Article Title]{Supplementary Information: Electrical control of valley polarized charged biexcitons in monolayer WS$_2$}


\author*[]{\fnm{Sarthak} \sur{Das}}\email{ds.sarthak.92@gmail.com}
\author[]{\fnm{Ding} \sur{Huang}}
\author[]{\fnm{Ivan} \sur{Verzhbitskiy}}
\author[]{\fnm{Zi-En} \sur{Ooi}}
\author[]{\fnm{Chit Siong} \sur{Lau}}
\author[]{\fnm{Rainer} \sur{Lee}}
\author[]{\fnm{Calvin Pei Yu} \sur{Wong}}
\author*[]{\fnm{Kuan Eng Johnson} \sur{Goh}}\email{kejgoh@yahoo.com}



\affil[]{\orgdiv{Institute of Materials Research and
Engineering (IMRE)}, \orgname{Agency for Science, Technology and
Research (A*STAR)}, \orgaddress{\city{Singapore}, \postcode{138634}, \country{Republic of Singapore}}}










\maketitle


\section*{Methods}\label{sec1}
\subsection*{Device Fabrication}
The prototype device was fabricated by sequential transfer of individual layers identified by optical microscope via dry transfer method using micro manipulators. The layers were exfoliated from commercially available flux-grown materials (WS$_2$ from 2D semiconductors, and graphene from HQ Graphene) on the top of polydimethylsiloxane (PDMS) using Nitto tape and transferred on pre-patterned contacts over a SiO$_2$(285 nm)/Si substrate. The contacts were defined through e-beam lithography using PMMA resist. The contact pads were made with 5/40 nm Cr/Au deposited by thermal evaporation followed by liftoff in Microposit remover 1165 solution. The entire stack was annealed inside a vacuum chamber ($10^{-6}$ mbar) at 250 $^{\circ}$C for 6 hours for better adhesion between the layers.
\subsection*{Measurement setup}
\subsubsection*{Polarized $\mu$PL}
For low-temperature circular dichroic photoluminescence (CDPL), the sample was placed inside a cryostat on top of a motorized stage. The excitation source for all-optical spectroscopy experiments was an NKT supercontinuum source (repetition rate 20 MHz, pulse duration ~80 ps for 5 nm bandwidth), filtered through a 770 nm shortpass filter and linearly polarized with a polarizing cube beam splitter. To create non-resonant excitation at 570 nm, the white supercontinuum was subsequently filtered through a 6 nm bandpass filter centred about 568 nm. The filtered beam was then circularly polarized through an achromatic $\lambda/4$ retarder and focused onto the sample through a $\times$50, NA = 0.6 microscope objective lens. The spot size on the sample is about 3 $\mu$m in diameter.  

Sample photoluminescence was captured through the same objective lens and converted from circular to linear polarization basis through the same achromatic $\lambda/4$ retarder. Excitation light was removed by passing the emission beam through a long pass filter. The emission beam was then split into two orthogonally polarized beams using a polarizing beam splitter and separately focused into two multimode fibers. The fibre outputs were then dispersed by an Acton SP300 spectrograph and simultaneously detected on non-overlapping regions of a Princeton Instruments EMCCD camera.
\subsubsection*{PLE measurements}
In resonant excitation and PLE experiments, the white supercontinuum beam was passed through a home-built prism monochromator, where the light was dispersed by an equilateral glass prism and then focused by a lens onto a pinhole. The focusing lens was mounted on a motorized stage to select wavelengths in the experimental range (580 – 600 nm). The selected wavelength had a bandwidth of $\sim$ 1 nm and a maximum power of $\sim$ 10 $\mu$W. For PLE, the emitted photons are passed through a 600 nm long pass filter before sending it to the spectrometer.   
\subsubsection*{Gate dependent spectroscopy}
For gate-dependent spectroscopy, we have used Keithley 2450 SMU as a source meter. The sample placed inside the cryostat was mounted to the chip carrier with wire bonding and connection to the SMU. The gate leakage current was below tens of pA for the different sets of measurements. For gate-dependent resonant excitation spectroscopy with 596 nm, the color line selected from the PLE path was passed through another 600/10 nm bandpass filter. 
We have estimated the carrier density from the following relation as $nq = CV_g$ where $n$ is the carrier density, $q$ is the elementary charge, $C$ is the effective capacitance and $V_g $ is the applied gate voltage. $C$ is calculated further with the relation as $C = [\epsilon_r \epsilon_0 / d ]|V_g|$. Here, $\epsilon_0$ =  8.85 $\times$ 10$^{-12}$ F.m$^{-1}$ is the permittivity of free space, $\epsilon_r$ = 3.9 \cite{wen2020dielectric,biswas2023rydberg} is the $h$BN dielectric constant and $d$ = 40 nm is the total thickness of the $h$BN layers.   

\newpage
\section*{Supporting Information S1}\label{sec2}
The figure below shows the optical image of the device fabricated using the dry transfer technique. The multi-peak fittings to examine the power-law for each energy state have been performed. The energy states remain nearly consistent regardless of the optical power applied.

\begin{figure}[h]%
\centering
\includegraphics[width=1.05\textwidth]{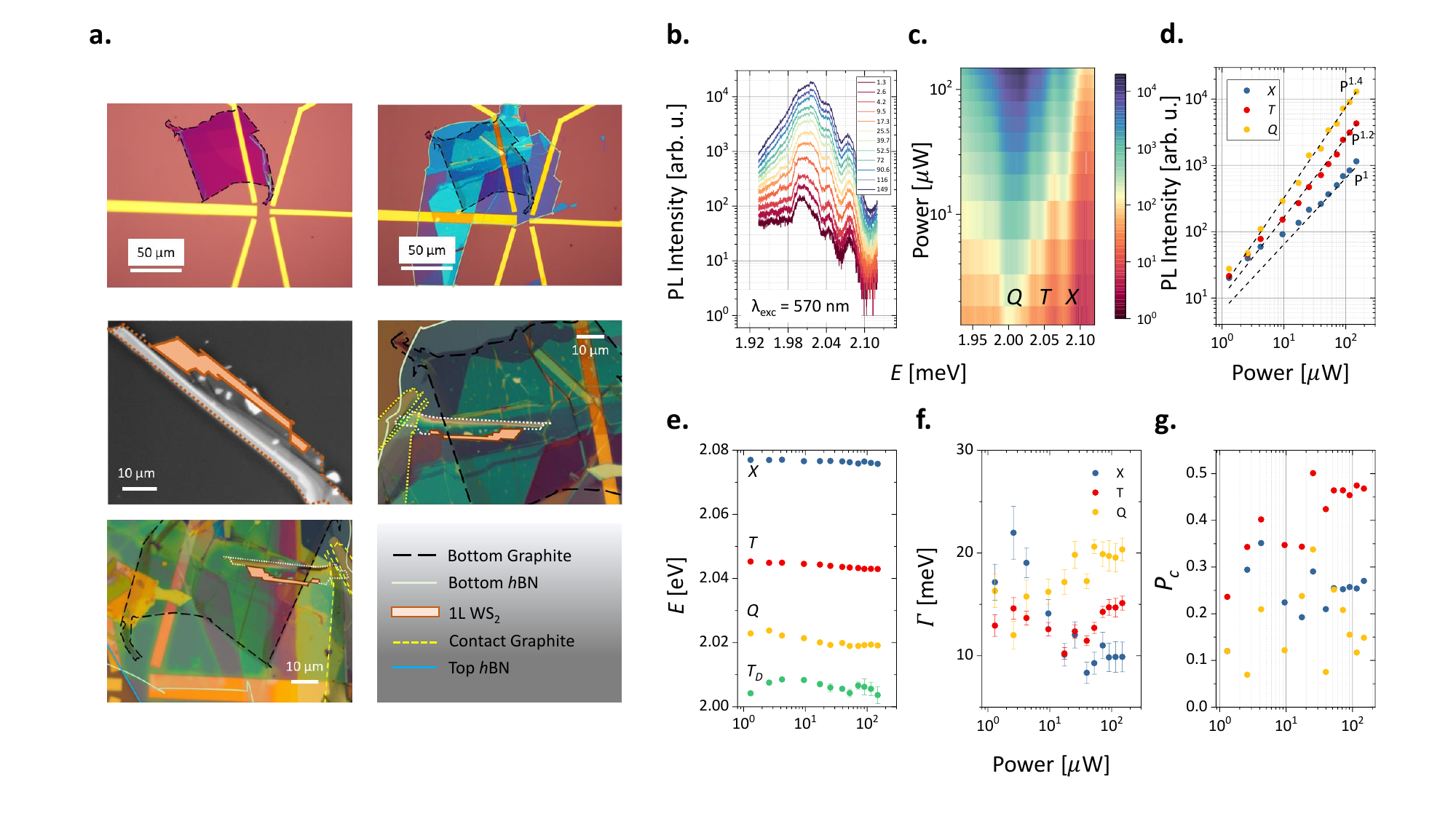}
\caption{\textbf{PL characterization of the device.} 
(a) Optical images of the device under test at different stages of the dry transfer method. Individual layers are outlined for reference.
(b-c) Power dependent $\mu$PL spectra for the monolayer WS$_2$ with non-resonant 570 nm excitation at $T$ = 4.6 K shown in semi-log scale and the 2D color plot of the same is presented in Fig. \ref{SF1}c. The labels are shown for reference on the color plot.
(d) The log-log plot of the power-dependent intensity of different peaks extracted from individual fittings. While the $X$ shows a linear increment with incident power, the charged biexcitons or the quinton state ($Q$) shows a super-linear behavior. 
(e-g) Different peak positions as a function of incident power show a negligible change over the operational range in Fig. \ref{SF1}e. The corresponding FWHM ($\Gamma$) of different peaks and the $P_c$ over the incident power is presented in Fig. \ref{SF1}f and Fig. \ref{SF1}g. 
}\label{SF1}
\end{figure}

\newpage
\section*{Supporting Information S2}\label{sec3}
Under the applied voltage range, the neutral and charged excitons display contrasting modulations. The charged states demonstrate a significant red shift as external doping is applied, as well as changes in intensity. The lower energy peaks (630-640 nm) experience a parallel red shift, similar to the $T_D$ states.  

\begin{figure}[h]%
\centering
\includegraphics[width=1.05\textwidth]{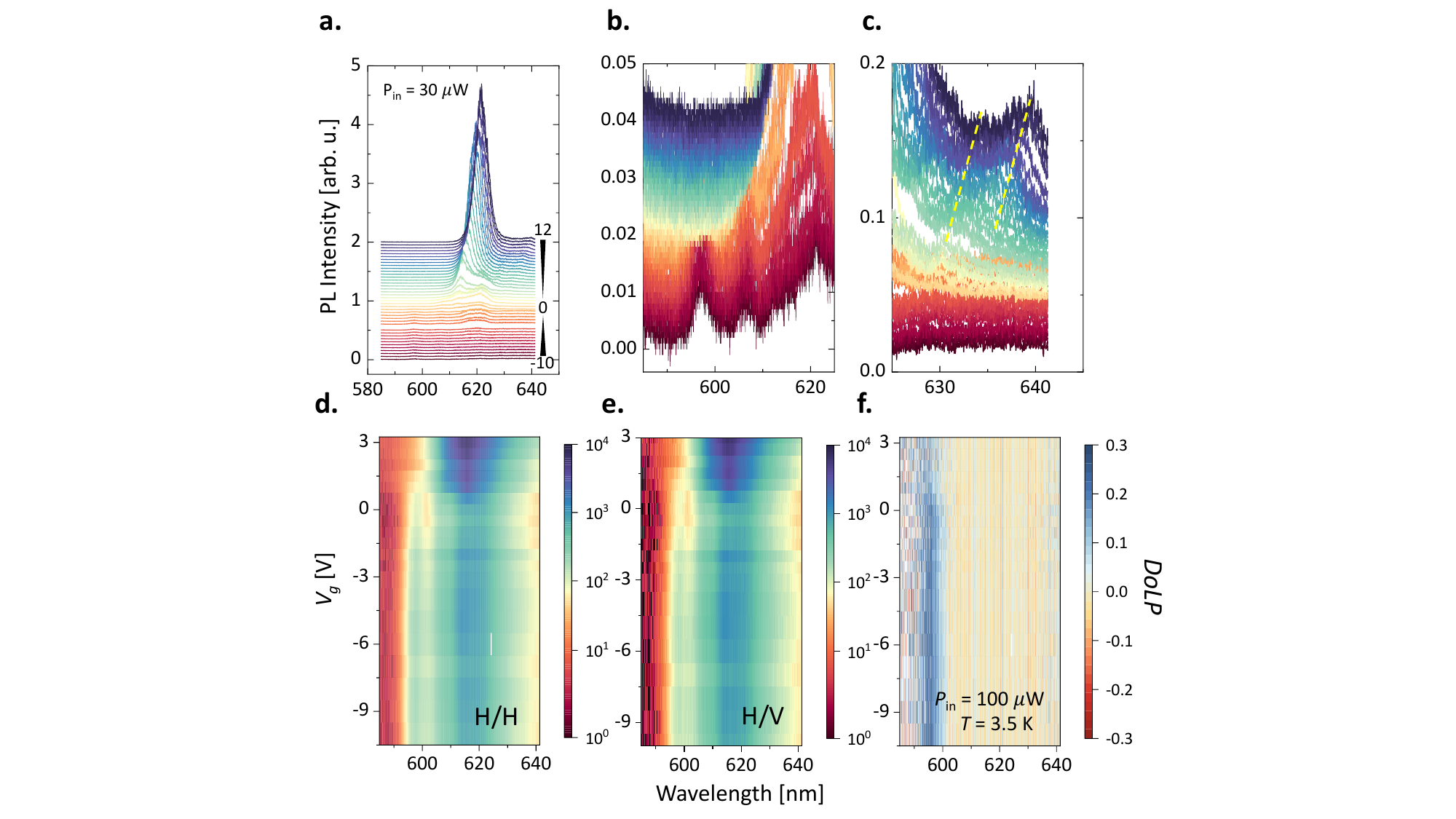}
\caption{\textbf{Gate dependent spectroscopy.} 
(a) The individual PL spectra under different gate voltages, $V_g$ = -10 V to 12 V, are discussed in the main text.
(b-c) The zoomed-in version of the spectra for low wavelength (high wavelength) in Fig. \ref{SF2}b(c). While the $X$ state only shows the intensity modulation, other states (charged states) are undergoing both the intensity variation and spectral shift. The peaks shown in \ref{SF2}c are referred to as the $T_D$ and their phonon replicas recently reported for WS$_2$ \cite{zinkiewicz2021excitonic, chand2023interaction}. 
(d-e) The gate-dependent color plot of the PL with linearly polarized light under parallel (H/H) and orthogonal (H/V) mode. 
(f) The color plot presents the degree of linear polarization (DoLP) for the spectra presented in \ref{SF2}d-e. Only the $X$ state shows linear polarization of $\sim$ 30 $\%$ which eventually decreases as the $X$ fades away for $V_g > 0$ V. 
}\label{SF2}
\end{figure}

\newpage
\section*{Supporting Information S3}\label{sec4}
The modulation of the $P_c$ and the linewidth ($\Gamma$) of the charged state with the PLE under $V_g = 0$ V. The linewidth is reduced for the charged states when the excitation is in resonance with the neutral exciton ($\lambda_{exc}$ = 596 nm). This trend is consistent for linearly polarized excitation also. 

\begin{figure}[h]%
\centering
\includegraphics[width=1.05\textwidth]{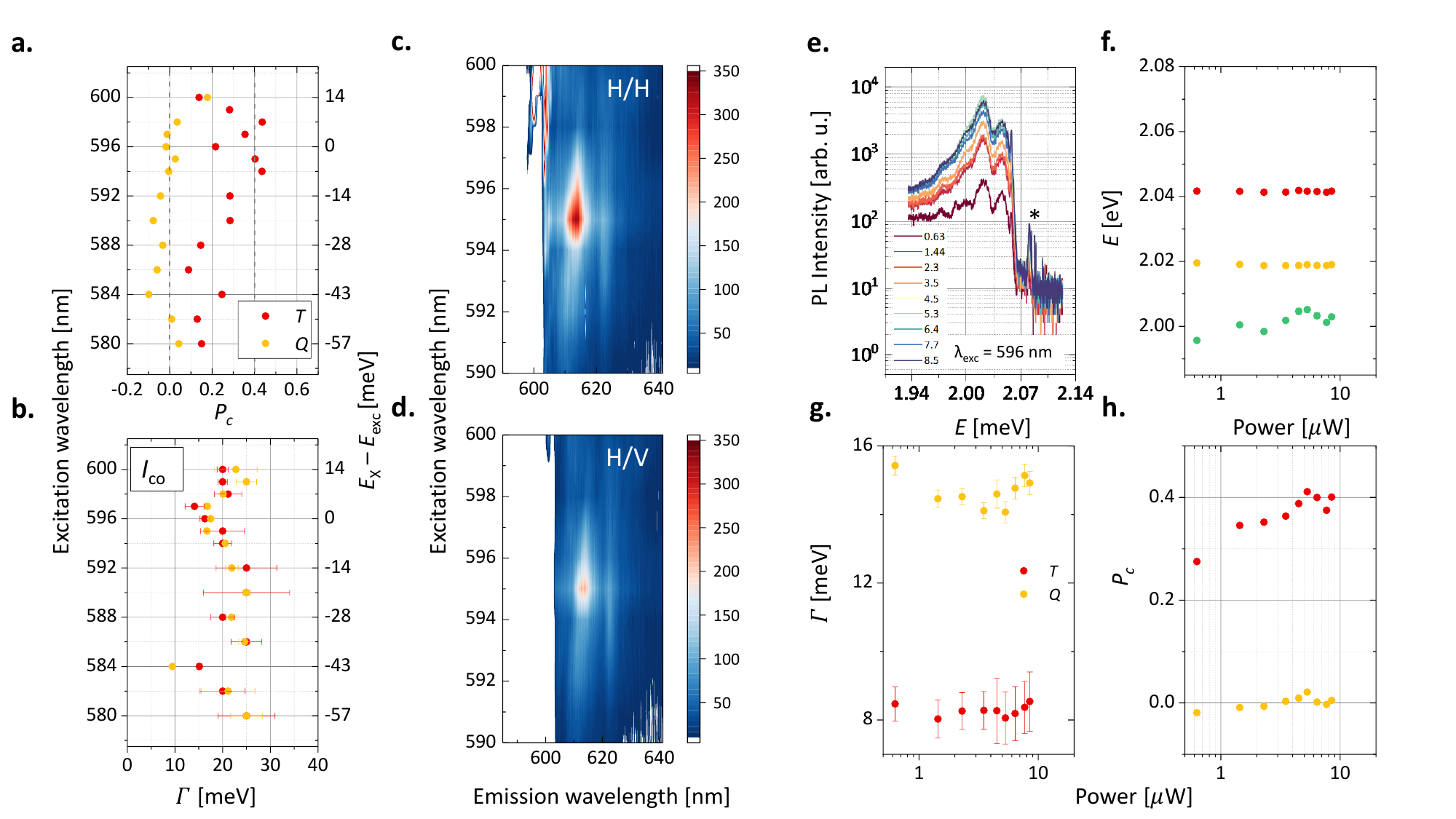}
\caption{\textbf{PLE characterization of the device and PL under resonant excitation.} 
(a-b) The modulation of valley polarization of the three and five particle states with excitation wavelength under $V_g = 0$ V. The $P_{c}$ is maximum for the trions when excess energy is minimum, while the $Q$ does not show valley polarization when $V_g = 0$ V. The FWHM ($\Gamma$) of the corresponding states as a function of excitation wavelength. 
(c-d) The PLE colorplot under linear polarized excitation with $V_g = 0$ V.
(e) The power-dependent PL spectra from monolayer WS$_2$ under exciton resonance ($\lambda_{exc}$ = 596 nm). The star(*) denotes the laser line.
(f-h) The extracted energy positions, FWHM, and the valley polarization as a function of incident power under resonance for individual charged energy states.
}\label{SF3}
\end{figure}

\newpage
\section*{Supporting Information S4}\label{sec5}
\begin{figure}[h]%
\centering
\includegraphics[width=1.05\textwidth]{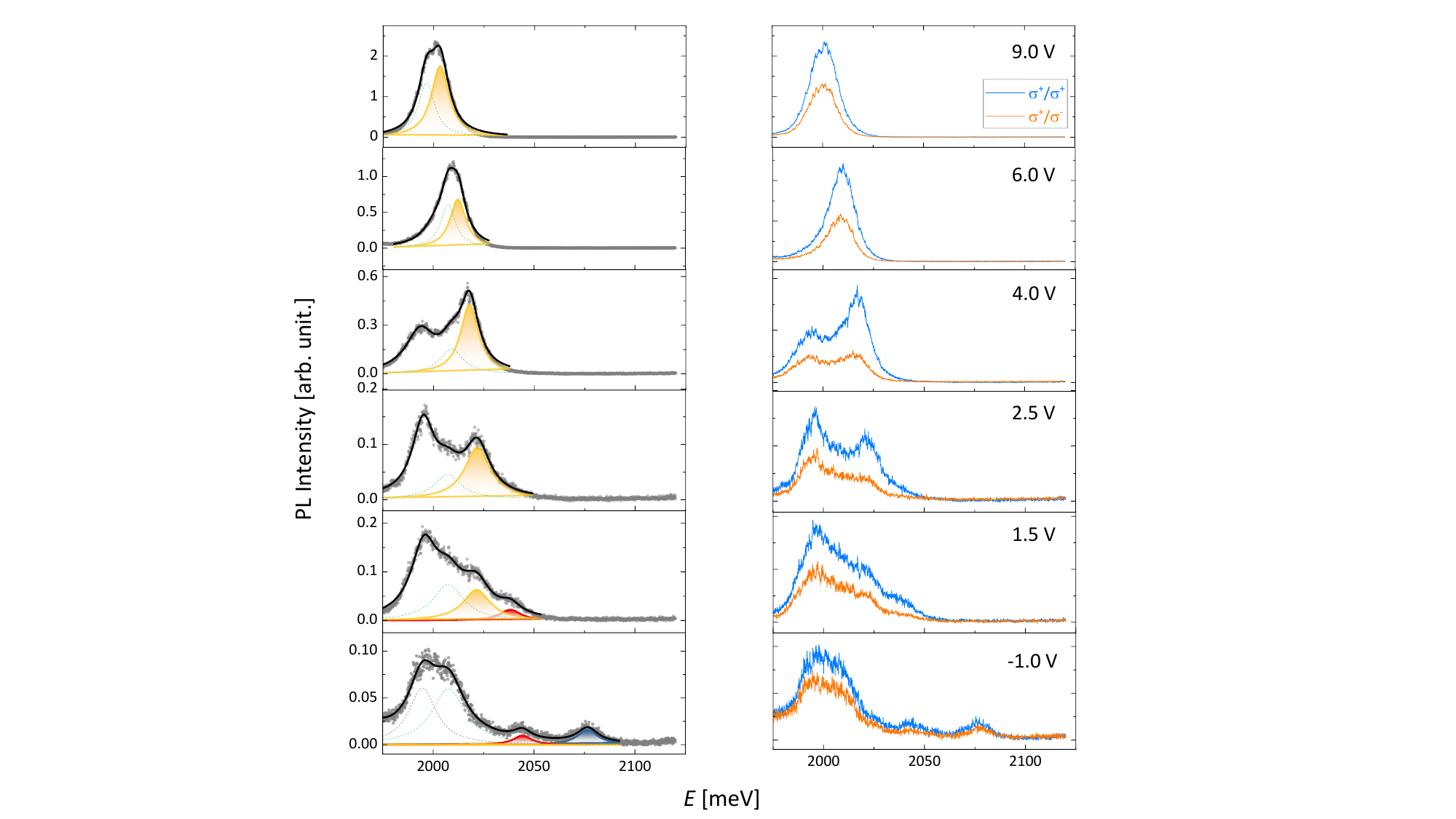}
\caption{\textbf{Sample peak fittings for non resonant excitation at $\lambda_{exc}$ = 570 nm.} 
Sample peak fittings for different gate voltages (discussed in the main text for 570 nm non-resonant excitation) where the $Q$ state is highlighted in yellow for reference showing the intensity modulation and spectral shift in the left panel. The individual spectra for given $ V_g$ under $\sigma^+/\sigma^+$ ($\sigma^+/\sigma^-$) excitation in blue (orange) is presented in the right panel.
}\label{SF4}
\end{figure}

\newpage
\section*{Supporting Information S5}\label{sec6}
The gate-dependent color plot for the non-resonant excitation ($\lambda_{exc}$ = 570 nm) under cross-polarized mode recorded $in-situ$ with the co-polarized mode, is presented below. The energy states behave similarly to the co-polarized mode. 

\begin{figure}[h]%
\centering
\includegraphics[width=1.05\textwidth]{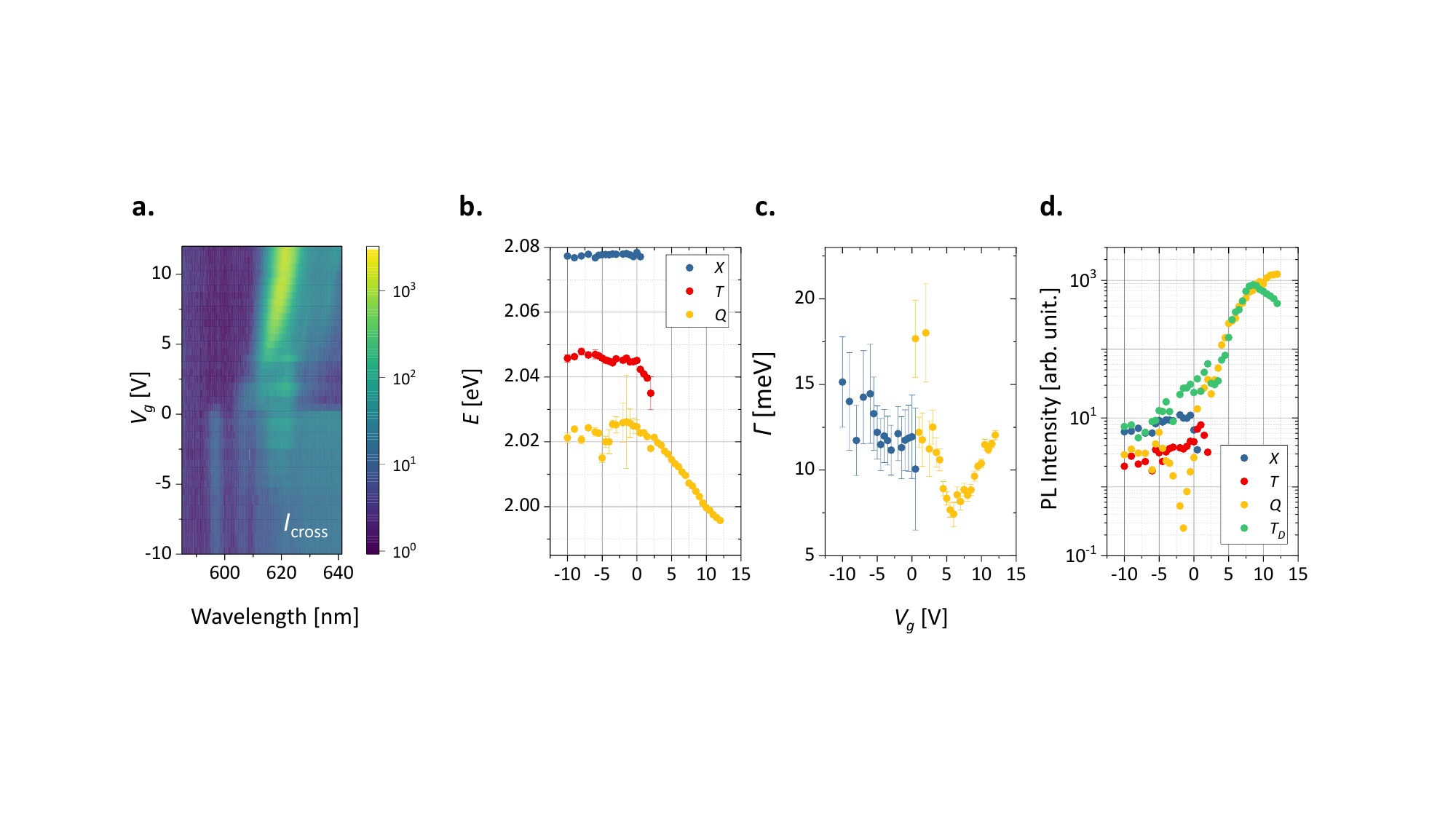}
\caption{\textbf{Gate dependent modulation of $\mu$PL with non-resonant excitation.} 
(a) The two-dimensional colorplot of the gate-dependent PL spectra with cross-polarized mode (co-polarized intensity is presented in the main text).
(b) The spectral position of the energy states extracted from Fig. \ref{SF5}a as a function of $V_g$.
(c) The modulation of FWHM of the $Q$ state with $V_g$ while the FWHM for the $X$ state remains unaltered for $V_g < 0$ V.
(d) The intensity modulation of the energy states over the applied $V_g$ extracted from \ref{SF5}a. The charged states are more responsive towards external bias and the brightness increases as the $n$-doping increases. 
}\label{SF5}
\end{figure}

\newpage
\section*{Supporting Information S6}\label{sec7}
Equations (2) - (4) from the main text provide the exchange interaction of the charged states over the q-space. It is shown that the valley polarization ($\mathcal{P_C}$) decreases as it moves away from the zone corners. Additionally, $\mathcal{P_C}$ depends on the ratio, $r$, of the radiative lifetime of the state to its scattering rate. To illustrate this, the $\mathcal{P_C}$ of the $Q$ state for non-resonant excitation from the experimentally mapped $q$-space (from figure 2a-b) with changing carrier density is plotted in relation to the radiative lifetime. 

\begin{figure}[h]%
\centering
\includegraphics[width=1.05\textwidth]{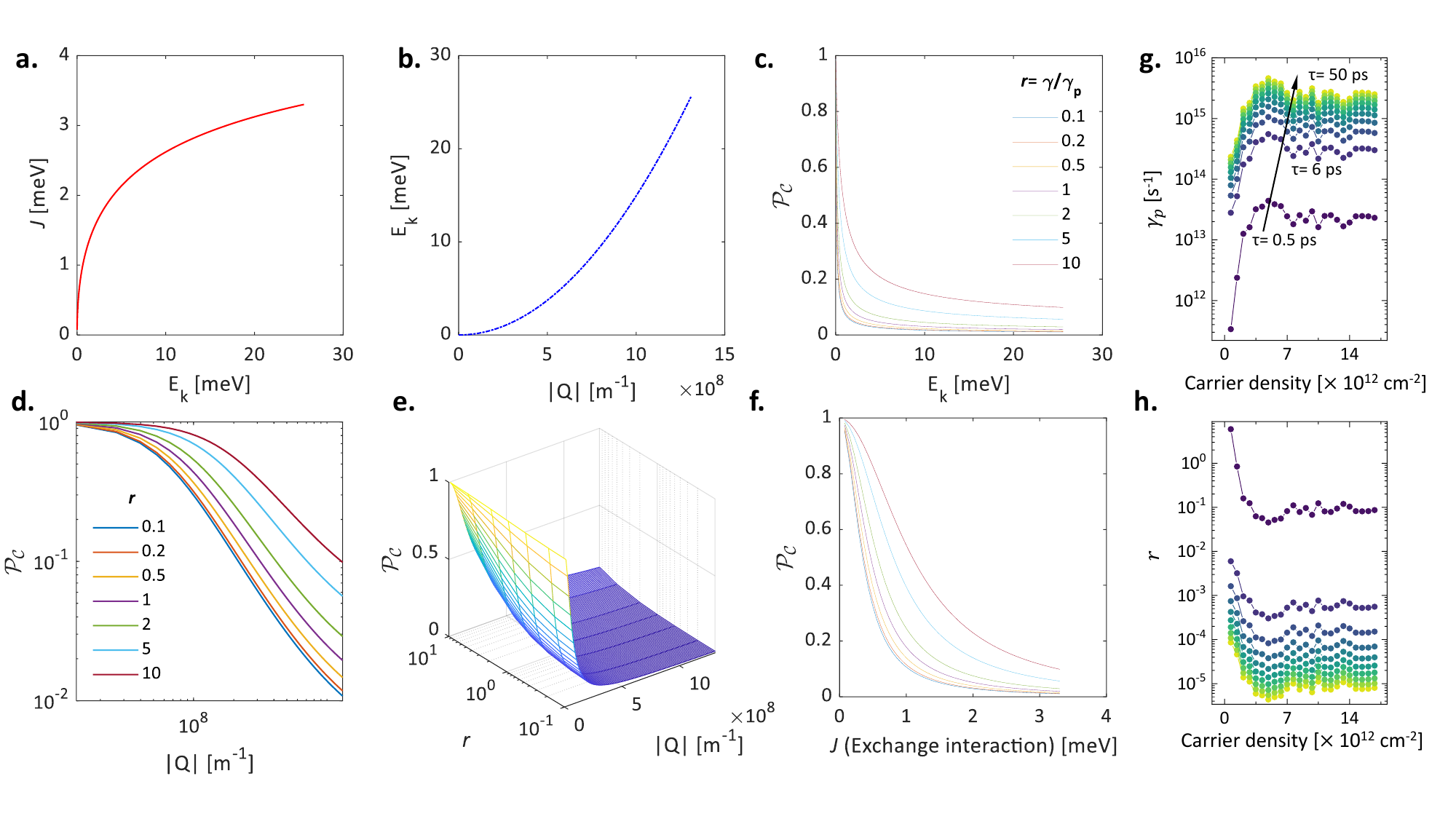}
\caption{\textbf{Exchange interaction ($J^{LR}_{e-h} (q)$) dependent modulation of valley polarization( $\mathcal{P_C}$)  of the charged states.} 
(a) Plot of exchange interaction \cite{chen2018superior, wu2021enhancement} variation as a function of increasing kinetic energy ($E_k$). The $E_k$ is relatively small for the excitons as their radiative recombination is limited within the light cone. Hence, the $J^{LR} (q)$ is small and approximated as constant for neutral states, which is a stark difference compared to the charged states \cite{wang2016radiative}. 
(b) The $E_k$ dispersion for the five particle states in $q$-space. The range of $E_k$ is limited to match the experimental redshift.  
(c-d) With increasing $E_k$ the $\mathcal{P_C}$ decreases as the $q$-space dispersion increases (plotted in Fig. \ref{SF6}d). However, the valley polarization is also dependent upon the ratio ($r$) between the decay rate ($\gamma$) and the scattering rate ($\gamma_p$).  
(e-f) The overall picture of the $\mathcal{P_C}$ modulation as a function of $r$ and $q$-dispersion (in Fig. \ref{SF6}e) as well as with the exchange interaction in Fig. \ref{SF6}f.
(g) The extracted scattering rate $\gamma_p$ from the experimental $\mathcal{P_C}$ value across the carrier density for varying radiative recombination time $\tau$. The variation in $\gamma_p$ across the carrier density with $\tau$ = 50 ps is presented in the main text.
(h) The decay rate normalized by the momentum scattering rate, $r$, ($r$ = $\frac{\gamma}{\gamma_p}$) against the carrier density for varying $\tau$. Note that if the $\tau$ changes with $V_g$ (or carrier density), the $r$ can be adjusted accordingly to match the experimental $\mathcal{P_C}$.
}\label{SF6}
\end{figure}

\newpage
\section*{Supporting Information S7}\label{sec8}
The PLE is performed when the monolayer is electron-doped in order to demonstrate the disappearance of the coherent coupling between neutral and charged states. The prominent dependence of the charged states exactly at excitonic resonance is missing at $V_g = +4$ V as presented below. 

\begin{figure}[h]%
\centering
\includegraphics[width=1.05\textwidth]{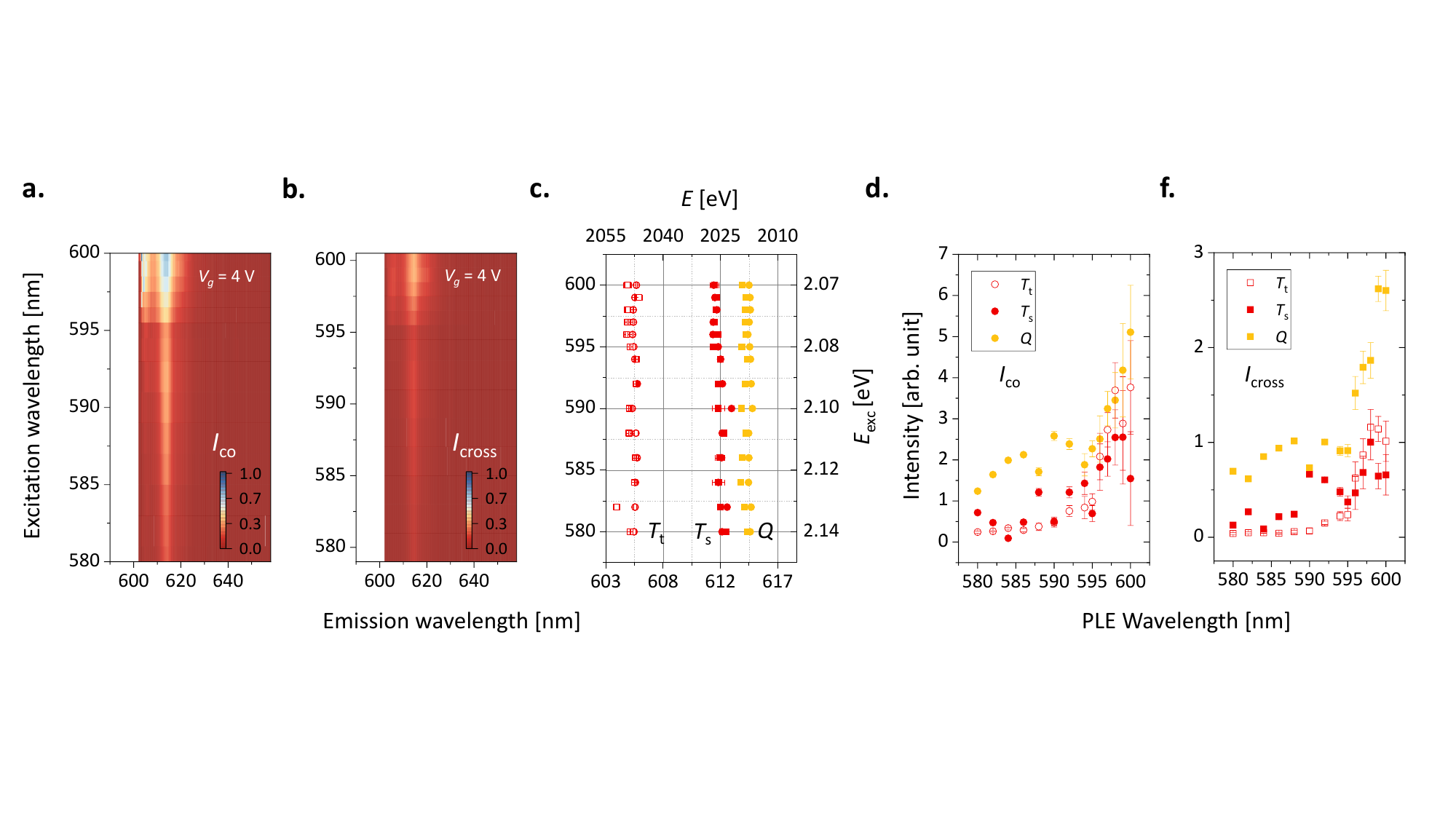}
\caption{\textbf{PLE characterization of the device in a $e^-$-rich environment.} 
(a-b) Colorplot of excitation wavelength-dependent circularly polarized emission for the charged states in an electron-rich environment with a constant $V_g = 4$ V. The charged states are not coherently coupled to the neutral states in contrary to the figure presented in the main text. 
(c) Wavelength-dependent energy positions of the charged states. There both the intra ($T_s$) and inter valley ($T_t$) trions are visible in the $n$-doped region. The circle (square) symbols denote the co (cross) polarized mode extracted from the fittings. 
(d-f) PLE wavelength-dependent intensity of the individual states shows interruption of the coherent coupling between neutral and charged states for both polarization as there is no prominent signature at exciton resonance. 
}\label{SF7}
\end{figure}

\newpage
\section*{Supporting Information S8}\label{sec9}
The gate-dependent color plot for the resonant excitation ($\lambda_{exc}$ = 596 nm) under cross-polarized ($\sigma^+/\sigma^-$) mode recorded $in-situ$ with the co-polarized mode, is presented below. The energy states behave similarly to the co-polarized mode along with a strong modulation in valley polarization similar to non-resonant excitation. 

\begin{figure}[h]%
\centering
\includegraphics[width=1.1\textwidth]{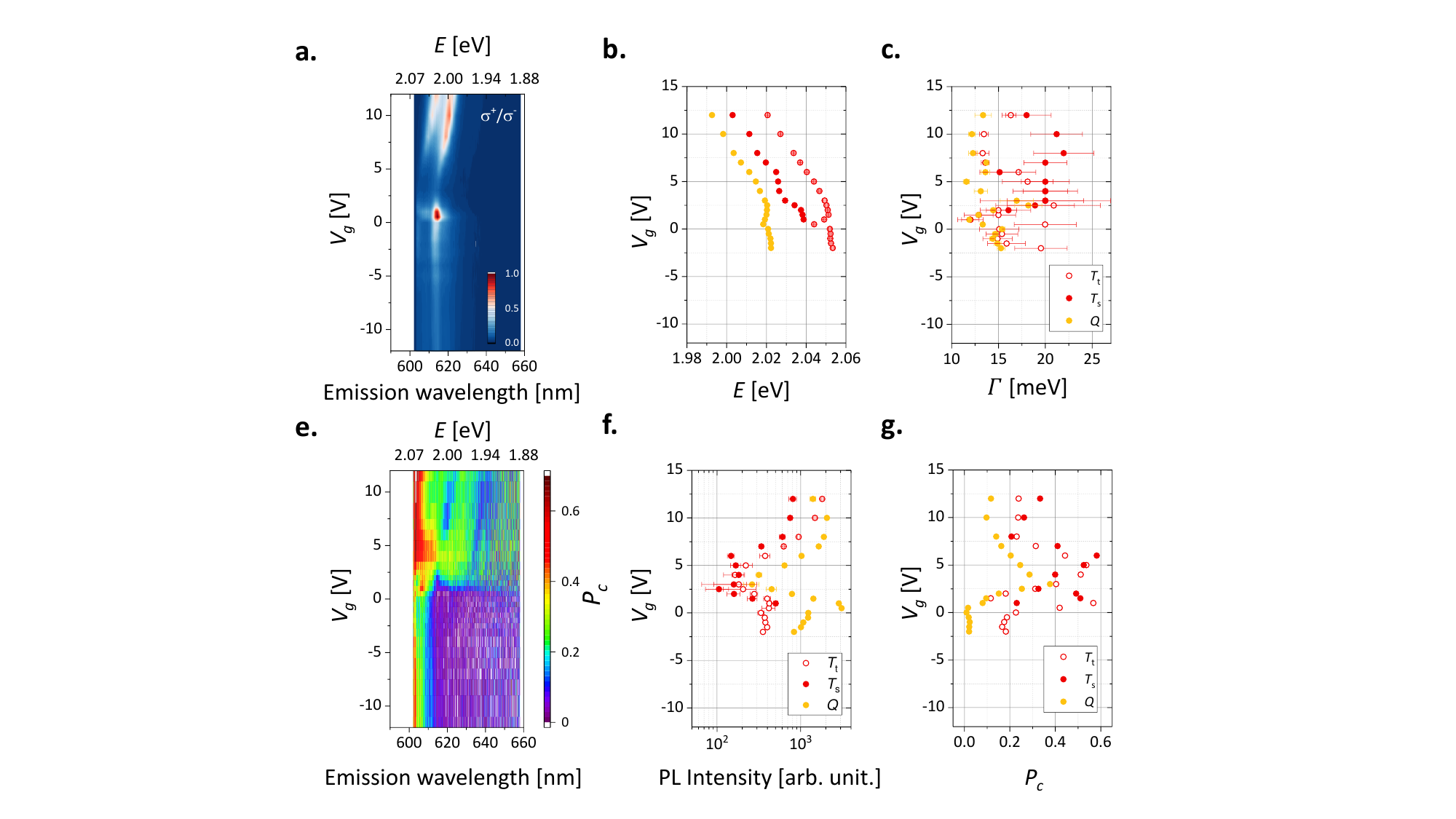}
\caption{\textbf{Gate dependent spectroscopy for resonant excitation.} 
(a) 2D colorplot of the $V_g$ dependent PL for $\lambda_{exc}$ = 596 nm recorded for $\sigma^+/\sigma^-$ emission mode. 
(b-c) $V_g$ dependent energy positions and corresponding line-width ($\Gamma$) for the charged energy states for $\sigma^+/\sigma^-$ emission mode.
(e) The calculated colormap for $V_g$ dependent degree of circular polarization for resonant excitation. While it is almost featureless for $V_g < 0$ V, the valley polarization shows a similar modulation to non-resonant excitation described in the main text for $V_g > 0$ V. 
(f-g) The anomalous modulation of the brightness of the charged states across the applied $V_g$ while the modulation of the valley polarization is presented in Fig. \ref{SF9}g.
}\label{SF8}
\end{figure}

\newpage
\section*{Supporting Information S9}\label{sec10}
\begin{figure}[h]%
\centering
\includegraphics[width=1.05\textwidth]{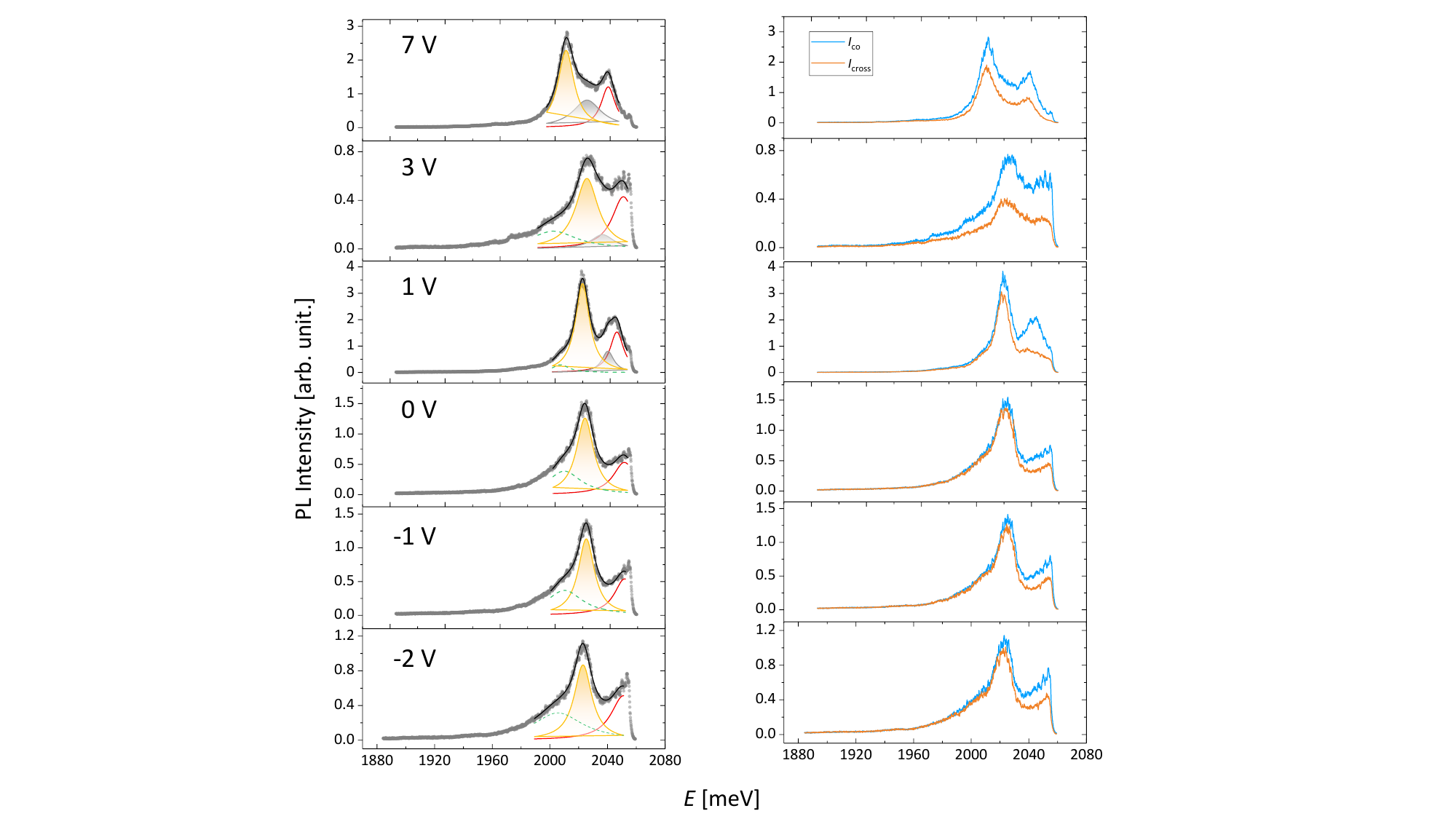}
\caption{\textbf{Sample peak fittings for resonant excitation, $\lambda_{exc}$ = 596 nm.} 
Sample peak fittings for different gate voltages (discussed in main text for 596 nm resonant excitation) where the $Q$ state is highlighted in yellow for reference showing the intensity modulation and spectral shift in left panel. The individual spectra for given $ V_g$ under $\sigma^+/\sigma^+$ ($\sigma^+/\sigma^-$) excitation in blue (orange) is presented in the right panel.
}\label{SF9}
\end{figure}

\newpage
\section*{Supporting Information S10}\label{sec11}
\begin{figure}[h]%
\centering
\includegraphics[width=1.05\textwidth]{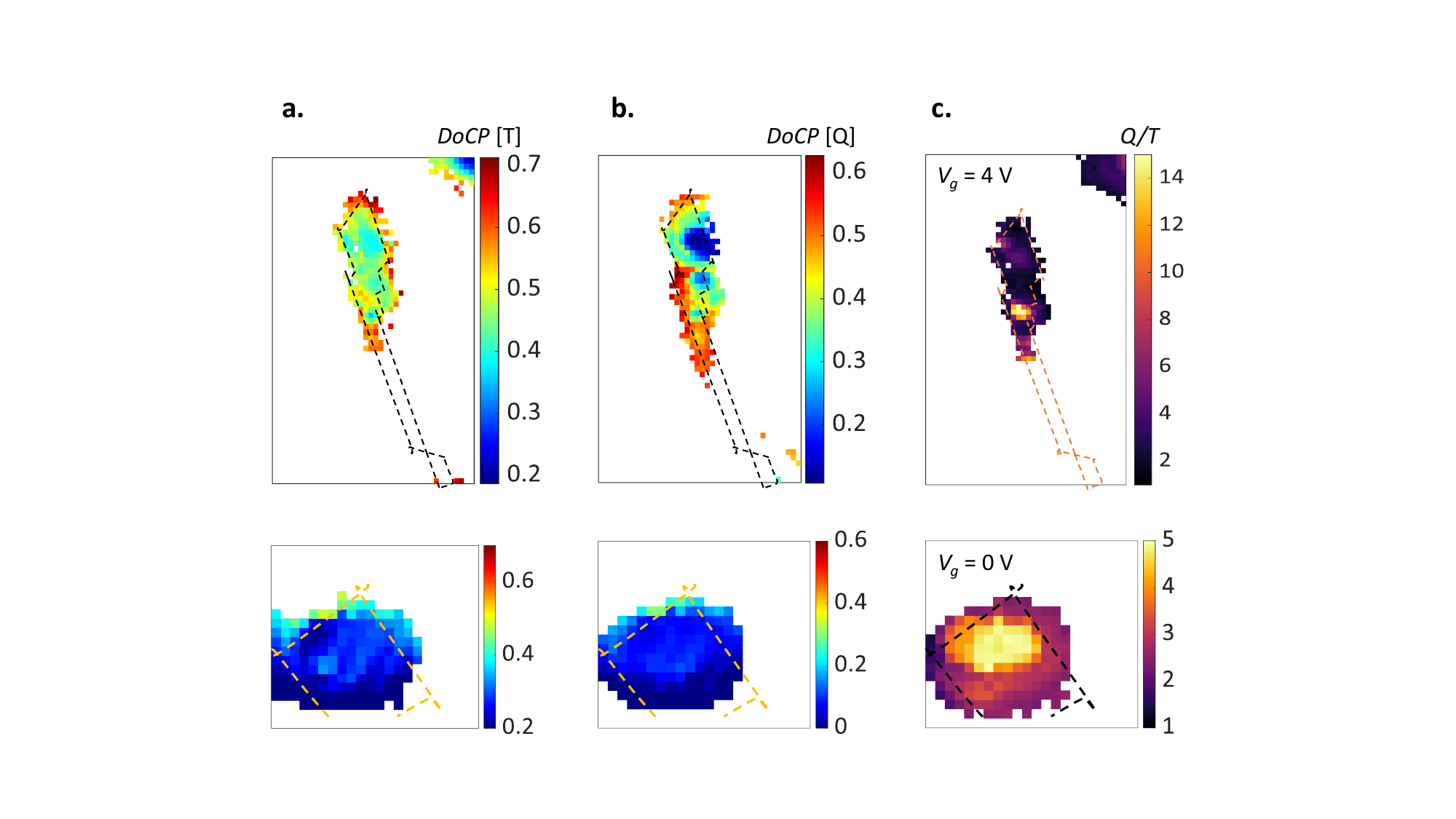}
\caption{\textbf{Spatial microscopy of the device for resonant excitation.} 
(a-b) The top panel shows the spatial microscopy of the device under test in an electron-rich environment (with a constant $V_g = 4$ V) showing the valley polarization within 606 - 610 nm region (corresponding to the three particle negatively charged excitons or trions) in a and the same for five-particle negatively charged biexciton (within 610 - 614 nm) in b. The monolayer region is outlined for reference. The signal-to-noise ratio is kept $\sim 10$ to remove the unwanted points. The bottom panel is for $ V_g = 0$ V for the same excitation.
(c) The spatial distribution of the intensity ratio between $Q$ and $T$ states showing the partial spatial inhomogeneity from the sample under test conditions. Thus, the gate-dependent $\mu$PL also has some dissimilar appearance over different locations within the same monolayer. The top panel is for $ V_g = 4$ V bottom panel is for $ V_g = 0$ V.
}\label{SF10}
\end{figure}

\newpage
\section*{Supporting Information S11}\label{sec12}
The gate-dependent PL spectroscopy is presented against the energy difference with respect to the laser excitation ($\lambda_{exc}$ = 596 nm) to correlate with the individual energy states with the available phonon modes.

\begin{figure}[h]%
\centering
\includegraphics[width=1.05\textwidth]{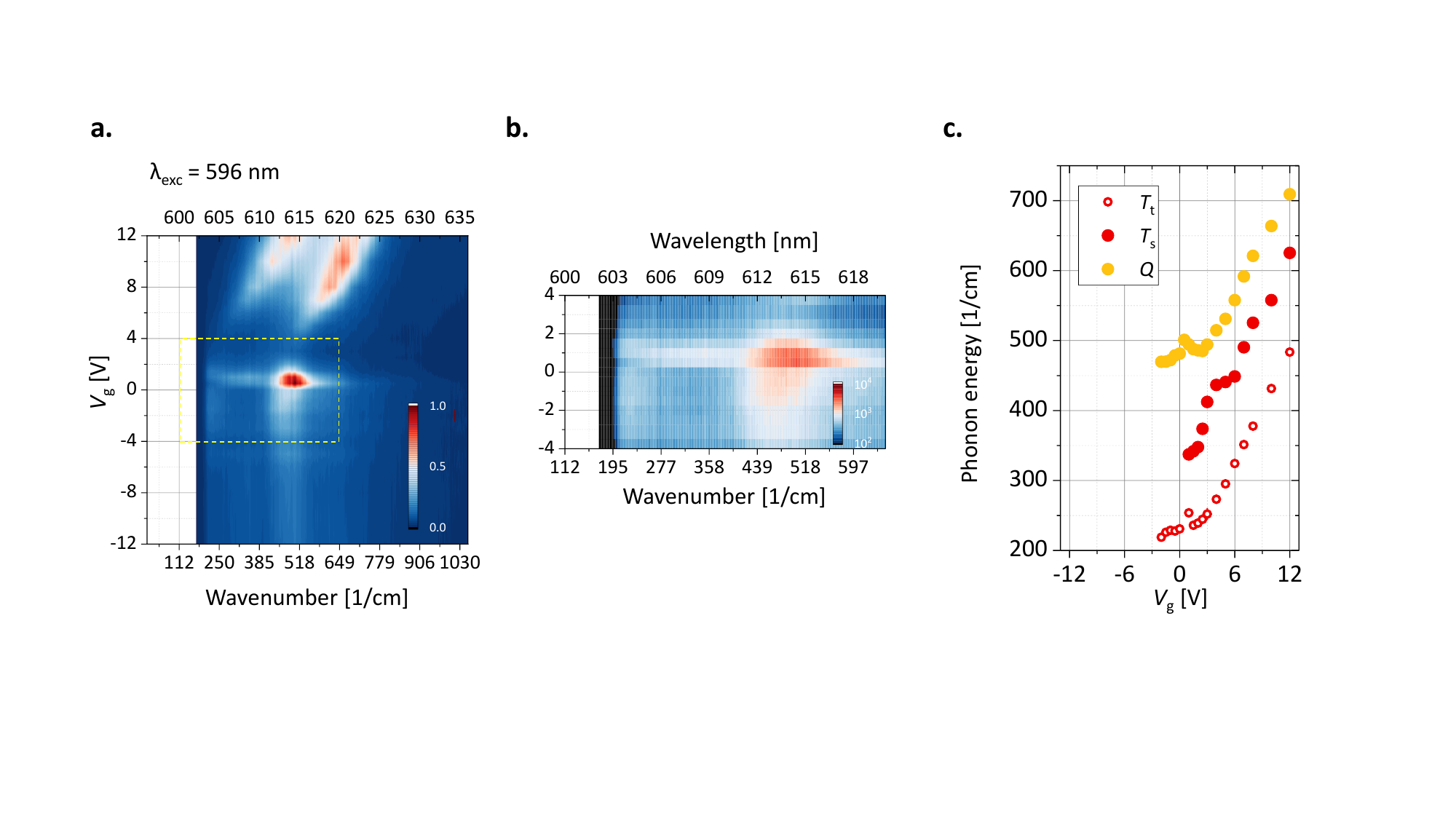}
\caption{\textbf{Exciton phonon interaction under resonant condition.} 
(a) The 2D colormap presented in the main text relative to the phonon-energy difference from the resonant excitation. 
(b) Zoomed in view of the outlined area showing that the brightness intensifies due to efficient down-conversion from the neutral state when the trion state matches the optical phonon energy at $V_g = 1$ V \cite{van2019virtual}.
(c) The shift of phonon energy difference from the excitation for different states over the applied $V_g$. While the energy difference of the three-particle state match with available phonon modes \cite{molas2017raman, van2019virtual}, the $Q$ state does not directly correlate to any of the phonon modes for monolayer WS$_2$.   
}\label{SF11}
\end{figure}

\section*{Supporting Information S12}\label{sec13}
\begin{figure}[h]%
\centering
\includegraphics[width=1.0\textwidth]{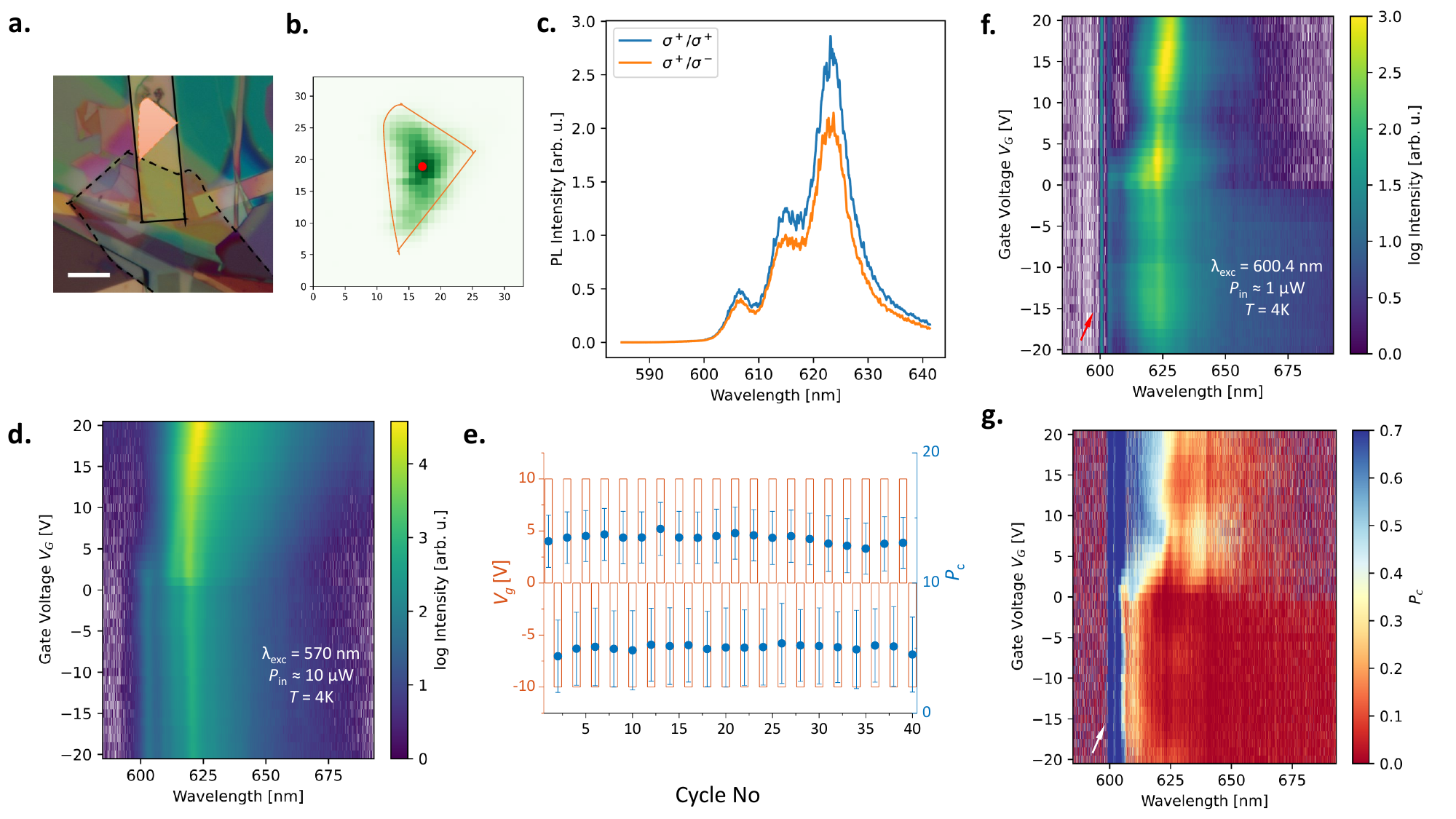}
\caption{\textbf{Gate-dependent spectroscopy and polarization switching for D2.} 
(a) The optical image of the device under test in the upper panel, where the monolayer portion is highlighted with a false-color. The back metal gate boundary is marked with a black line while the contact graphite is shown with a dashed-black boundary. The scale bar is 10 $\mu$m. 
(b) The two-dimensional PL mapping (within 16.5 $\mu$m $\times$ 16.5 $\mu$m area) of the luminescence intensity from the device within 585-645 nm wavelength range at $T$ = 4 K in co-polarized mode. The monolayer boundary is highlighted for reference. 
(c) The representative PL spectra for co- and cross-polarized mode from the map (the red dot) under 570 nm excitation with $<$10 $\mu$W excitation power ($P_{in}$), showing dominant features corresponding to different energy states.
(d) The gate-dependent PL spectroscopy for non-resonant ($\lambda_{exc}$ = 570 nm) excitation. The energy states produce similar observations discussed in the main text. However, due to the thicker dielectric, the carrier density is less compared to the main text device. Hence, the spectral red shift corresponding to the dominant $Q$ state is subtle for D2 compared to D1.
(e) Switching between high- and low-polarized states with altering $V_g$ for $\lambda_{exc}$ = 570 nm. As previously discussed, the modulation in $P_c$ is less pronounced in this case, and the applied voltage $V_g$ must be increased to $\pm 10$ V to achieve a noticeable switch in $P_c$ at a specific carrier density.
(f-g) 2D colorplot of the gate-dependent PL and $P_c$ for quasi-resonant ($\lambda_{exc}$ = 600.4 nm) excitation. The $P_c$ modulation here is remarkably higher, along with the anomalous intensity profile for the charged states against the carrier density. The arrow (red in f, and white in g) denotes the laser line.
}\label{SF12}
\end{figure}

To validate our experimental results discussed in the main text, we have performed gate-dependent PL spectroscopy on another device (D2) with a similar architecture. Here, the back gate is replaced by a metal gate (Cr/Au) in place of a graphite gate as shown in figure \ref{SF12}a. The device shows a similar performance as the previous device (D1) discussed in the main text. The exciton resonance ($\sim$ 605 nm) here is red-shifted compared to D1 likely due to different $h$BN thicknesses (in top and bottom) and the change in the binding energy \cite{gerber2018dependence}. Hence, for non-resonant excitation ($\lambda_{exc}$ = 570 nm), the modulation in $P_c$ is lower, likely due to excitation even further from the optical bandgap as shown in figure \ref{SF12}d-e. However, in quasi-resonant excitation ($\lambda_{exc}$ = 600.4 nm), the gate dependence of the PL emission replicates the non-monotonic evolution of the charge states under electron doping shown in figure \ref{SF12}f. Moreover, the $P_c$ can be modulated electrically (shown in figure \ref{SF12}g), as discussed for the device in the main text.

\newpage
\section*{Supporting Information S13}\label{sec14}
By altering the polarity of $V_g$, valley polarization can be achieved in either the on or off state by considering a threshold. By applying a sequential series of opposite-polarity pulses of voltage $V_g$, the degree of circular polarization ($P_c$) of the charged states can also be altered. This is applicable for both the non-resonant and resonant excitations as shown below. The average $P_c$ has been considered within a fixed wavelength window, ignoring the relative of the individual energy states over the applied $V_g > 0$ V (where the states do not show any spectral shift for $V_g < 0$ V). 

\begin{figure}[h]%
\centering
\includegraphics[width=1.05\textwidth]{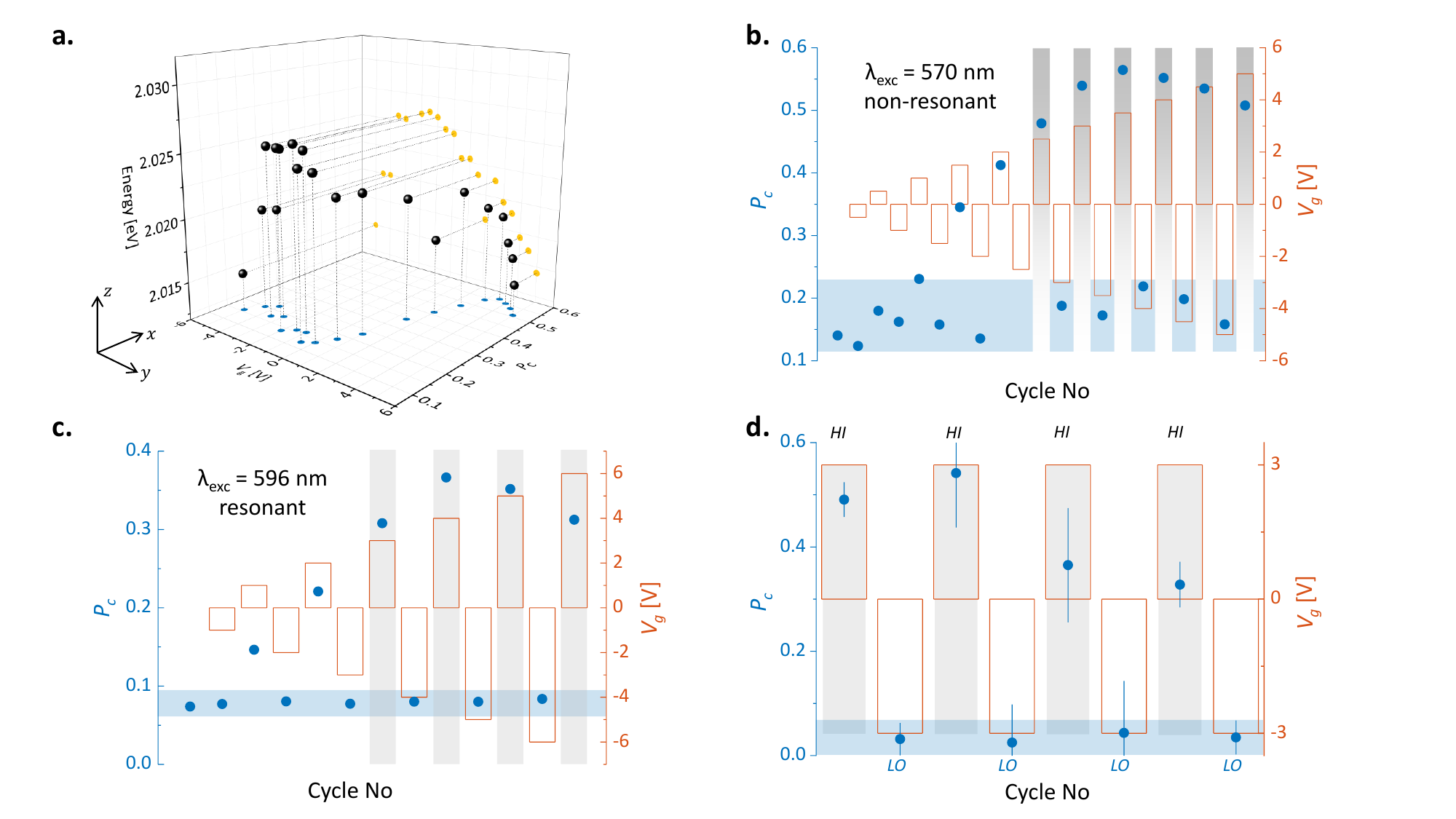}
\caption{\textbf{Switching of valley polarization for D1.} 
(a) The 3D plot of the $Q$ state under $V_g$ variation (in $y$-axis) against its energy position (in $z$-axis) and $P_c$ (in $x$-axis) from figure 2b-c from the main text under non-resonant excitation ($\lambda_{exc}$ = 570 nm). 
(b) When an increasing $V_g$ of alternating polarity ($-6$ V $< V_g <$ $+6$ V) is applied, the corresponding valley polarization also shows an alternating trend for the $Q$ state. Considering the $P_c$ within $0.1 - 0.2$ as low-polarized state (shown in the blue-shaded region) and $P_c > 0.4$ as high-polarized state for a wavelength window of 610-617 nm, the $P_c$ can be realized as a voltage-controlled valley switch for non-resonant excitation. 
(c) The switching of valley polarization for resonant excitation. Here, the threshold can be considered as $P_c < 0.1$ for $V_g < 0$ V. The $P_c$ from the individual states are slightly different as extracted from fitting. Here, the average $P_c$ has been considered within a fixed wavelength window (609-616 nm) with a primary contribution from the $Q$ state.
(d) When a fixed $V_g$ is applied with an alternative polarity, then also the $P_c$ can be switched repeatedly between a high (HI) and a low (LO)-polarized state. Here, it is shown for $V_g$ = $+3$ V and $-3$ V, with a fixed smaller wavelength window (614-616 nm). This can further be implemented for any voltage between $2 < |{V_g}| < 5$ V. This shows further deterministic control over photon energy along with voltage-controlled polarization tuning.
}\label{SF13}
\end{figure}


\newpage
\bibliographystyle{naturemag}
\bibliography{ref_sub}